\documentclass[preprint,emulateapj,twocolumn]{aastex}
\usepackage{epsf}
\usepackage{emulateapj5}
\usepackage{onecolfloat}

\newcommand{\hvol}{h^{3}{\mathrm{Mpc}}^{-3}}

\newcommand{\hkpc}{h^{-1}\mathrm{kpc}}
\newcommand{\hpc}{h^{-1}\mathrm{pc}}
\newcommand{\hMsun}{\ h^{-1}\mathrm{M}_{\odot}}
\newcommand{\hMpc}{\ h^{-1}\mathrm{Mpc}}

\newcommand{\kms}{{\,{\rm km}\,{\rm s}^{-1}}}
\newcommand{\kpc}{{\,{\rm kpc}}}
\newcommand{\Gyr}{{\,{\rm Gyr}}}

\newcommand{\Omegam}{\Omega_{M}}

\newcommand{\Omegal}{\Omega_{\Lambda}}
\newcommand{\rhomean}{\rho_{\mathrm{M}}}
\newcommand{\deltacrit}{\delta_{\mathrm{c}}}

\newcommand{\Fsurv}{F_{\mathrm{surv}}}
\newcommand{\fsat}{f_{\mathrm{sat}}}

\newcommand{\Nsg}{N_{\mathrm{sat}}}
\newcommand{\Nsat}{\left<N_{\mathrm{sat}}\right>}

\newcommand{\NNsat}{\left<N_{\mathrm{sat}}(N_{\mathrm{sat}}-1)\right>}
\newcommand{\Ntotal}{N_{\mathrm{total}}}

\newcommand{\vmax}{V_{\mathrm{max}}}
\newcommand{\vmaxsat}{V_{\mathrm{max}}^{\mathrm{sat}}}
\newcommand{\vmaxhost}{V_{\mathrm{max}}^{\mathrm{host}}}

\newcommand{\tlookback}{t_{\mathrm{lookback}}}
\newcommand{\tdyn}{t_{\mathrm{dyn}}}

\newcommand{\Msat}{M_{\mathrm{sat}}}
\newcommand{\Rsat}{R_{\mathrm{sat}}}

\newcommand{\Mhost}{M_{\mathrm{host}}}

\newcommand{\Mmin}{M_{\mathrm{min}}}
\newcommand{\Rvir}{R_{\mathrm{vir}}}
\newcommand{\Mvir}{M_{\mathrm{vir}}}
\newcommand{\rs}{r_{\mathrm{s}}}
\newcommand{\rt}{r_{\mathrm{t}}}

\newcommand{\rtide}{r_{\mathrm{tide}}}
\newcommand{\rmerge}{r_{\mathrm{merge}}}

\newcommand{\rmax}{r_{\mathrm{max}}}
\newcommand{\Vvir}{V_{\mathrm{vir}}}
\newcommand{\Rcirc}{R_{\mathrm{circ}}}
\newcommand{\Jcirc}{J_{\mathrm{circ}}}
\newcommand{\Vcirc}{V_{\mathrm{circ}}}
\newcommand{\Dvir}{\Delta_{\mathrm{vir}}}
\newcommand{\cvir}{c_{\mathrm{vir}}}

\newcommand{\lcdme}{\Lambda {\mathrm{CDM}}_{80}}

\newcommand{\ac}{a_{\mathrm{c}}}
\newcommand{\aobs}{a_{\mathrm{o}}}

\newcommand{\dd}{\mathrm{d}}

\newcommand{\beq}{\begin{equation}}
\newcommand{\eeq}{\end{equation}}

\newcommand{\lsim}{\lower0.6ex\vbox{\hbox{$ \buildrel{\textstyle <}\over{\sim}\ $}}}
\newcommand{\gsim}{\lower0.6ex\vbox{\hbox{$ \buildrel{\textstyle >}\over{\sim}\ $}}}

\bibpunct[,]{(}{)}{;}{a}{}{,}

\begin{document}

\slugcomment{{\em The Astrophysical Journal, submitted}}

\twocolumn[

\title{
The Physics of Galaxy Clustering I: 
A Model for Subhalo Populations
}

\lefthead{Zentner et al.}
\righthead{The Physics of Galaxy Clustering I}

\author{
Andrew R. Zentner\altaffilmark{1}, 
Andreas A. Berlind\altaffilmark{3}, 
James S. Bullock\altaffilmark{4,5,6}, \\
Andrey V. Kravtsov\altaffilmark{1,2}, 
and Risa H. Wechsler\altaffilmark{1,2,6,7}
}

\begin{abstract}
  
  We present a semi-analytic model for Cold Dark Matter halo 
  substructure that can be used as a framework for studying the
  physics of galaxy formation and as an ingredient in halo models of
  galaxy clustering.  The model has the following main ingredients:
  (1) extended Press-Schechter mass accretion histories; (2) 
  host halo density profiles computed according to the
  trends observed in cosmological simulations; (3) distributions 
  of initial orbital parameters of accreting subhalos measured in a 
  high-resolution simulation of three Milky Way-size halos; and 
  (4) integration of the orbital evolution of subhalos including
  the effects of dynamical friction and tidal mass loss.  We perform a
  comprehensive comparison of the model calculations to the results of 
  a suite of high-resolution cosmological simulations. The comparisons
  show that subhalo statistics, such as the velocity and mass
  functions, the radial distributions, and the halo occupation
  distributions agree well over three orders of magnitude in host halo
  mass and at various redshifts.  We find that both in
  the simulations and in our model the radial distributions of subhalos
  are significantly shallower than that of the dark matter density.  
  The abundance of
  subhalos in a host is set by competition 
  between tidal disruption and new accretion.  
  Halos of high mass and
  halos at high redshift tend to host more subhalos because the
  subhalos have, on average, been accreted more recently.  Similarly, 
  at a fixed mass and epoch, halos that formed more recently host a
  larger number of subhalos. Observed ``fossil groups'' may represent
  an extreme tail of this correlation. We find a related correlation
  between host halo concentration and satellite abundance at fixed
  host mass, $\Nsg \propto \cvir^{-a}$, where $a$ changes with
  redshift and host-to-subhalo mass ratio.  Lastly, we use our
  substructure model to populate host halos in one of the
  high-resolution cosmological simulations, replacing the actual
  subhalos resolved in this simulation and using the host mass as the
  only input for the model calculation.  We show that the resulting
  correlation function of such a hybrid halo ensemble is
  indistinguishable from that measured directly in the simulation.
  This supports one of the key tenets of the standard halo
  model --- the assumption that the halo occupation distribution is
  statistically independent of the host halo environment.
\end{abstract}

]

\altaffiltext{1}{Kavli Institute for Cosmological Physics and 
Department of Astronomy and Astrophysics, The University of Chicago, 
Chicago, IL 60637, USA; zentner,risa@kicp.uchicago.edu,andrey@oddjob.uchicago.edu}
\altaffiltext{2}{The Enrico Fermi Institute, The University of Chicago, 
Chicago, IL 60637, USA}
\altaffiltext{3}{Center for Cosmology and Particle Physics and 
Department of Physics, New York University, 
New York, NY 10003, USA; aberlind@cosmo.nyu.edu}
\altaffiltext{4}{Harvard-Smithsonian Center for Astrophysics, 
Cambridge, MA 12138, USA; jbullock@cfa.harvard.edu}
\altaffiltext{5}{Department of Physics and Astronomy, 
The University of California at Irvine, Irvine, CA 92697, USA}
\altaffiltext{6}{Hubble Fellow}
\altaffiltext{7}{Enrico Fermi Fellow}


%
%

\section{Introduction} \label{sec:intro}

Numerical simulations of structure formation set within the cold dark
matter (CDM) paradigm \citep[see ][]{white_rees78, blumenthal_etal84}
have revealed that virialized, dark matter halos are rife with
distinct, self-bound substructure or {\em subhalos}
\citep*[e.g.][]{moore_etal99, klypin_etal99, klypin_etal99b,
  ghigna_etal98,ghigna_etal00, kravtsov04b,diemand_etal04,
  gao_etal04a, reed_etal04}.  It is tempting to associate subhalos,
particularly large subhalos, with galaxies in groups and clusters
\citep{klypin_etal99,kravtsov_klypin99,colin_etal99,moore_etal99}.
Lending support to this conjecture, \citet{kravtsov04a} used natural
assumptions to connect galaxy luminosity to subhalos and showed that
the clustering properties of subhalos and their {\em host} halos match
the observed clustering properties of galaxies \citep[see
also][]{neyrinck_etal04}.  This implies that the occupation
distribution of subhalos must be quite similar to the occupation
distribution of galaxies in host halos \citep[e.g.][]{berlind02} and
that luminous satellite galaxies are likely associated with massive
dark matter subhalos.  A corollary of this is that if one understands
the physical processes that govern subhalo occupation, one gains
significant insight into the physics of galaxy clustering.

The abundances  of subhalos     within  host dark matter  halos    are
determined through the  interplay of several important and  relatively
distinct physical processes.    If one imagines,  for simplicity,  the
development of a  subhalo  population as a   sequence, the  first
process to  consider is the merging of halos to form ever larger 
systems.  In the prevailing CDM paradigm of hierarchical structure 
growth, halo merger rates are governed  by 
the initial density field and the global cosmological model, 
and tend to be more rapid at early times.  The second step is 
to consider the evolution of merged subhalos in the potential of 
the host halo.  In contrast to merger statistics, 
subhalo evolution is influenced by local processes in the 
high-density environment of the host, 
like dynamical friction, tidal  mass loss, 
and heating by  violent interactions with other structures.
These processes reduce the amount of distinct
substructure  within host systems over 
timescales of order the local dynamical time.
At early times, when  the  merger timescale is 
short compared to the  dynamical time, 
we expect high substructure counts within individual host 
halos.  At late times, we expect substructure to be
reduced, as subhalos are packed into more and more dense environments
inside larger and larger host systems.  All of these physical
processes are purely gravitational, but have a significant  effect on
the small-scale  clustering of subhalos.  By implication, galaxy
clustering statistics are governed by these processes as well.

This paper is the first of a  series  of studies of the physical 
processes that govern galaxy clustering statistics.    In the 
context of this series, this paper represents a description of 
our theoretical methodology, but the content of this paper 
is considerably more general.  We present a
semi-analytic model  that can  be  used  to  make predictions  for the
substructure populations of dark matter halos.   The model begins with
the computation  of the merger    histories of dark  matter halos  and
approximates the subsequent  effects of dynamical  friction and
tidal  mass  after  accretion.    In the  regime where   they are
commensurable,   we make detailed   comparisons  of this  model to the
results of high-resolution $N$-body simulations to assess the accuracy 
and applicability of the model (see \S~\ref{sec:results}).  
We emphasize that the model we present in this first paper 
includes only the effect of gravity and neglects any effects 
of baryon cooling and condensation.  Our strategy in so doing 
is to build upon a tractable model that can be robustly be 
tested with $N$-body simulations in a variety of ways.  
In this way, the importance of additional processes and 
approximations can be ascertained in a well-controlled 
way and we will pursue these aims in the subsequent papers 
of this series.

There are advantages to using an 
analytic model that has been tested against existing 
simulations in a wide range of applications.  
First, an analytic model  can quickly generate a 
statistically-large  number of realizations  of subhalo 
populations within host halos of various masses with a relatively 
small computational effort.  As a result, such a model can 
complement direct simulation where limited dynamic range 
severely restricts the size of any host halo sample in 
a study of halo substructure.  Along these lines, 
\citet{islam_etal03} used a semi-analytic approach 
to study a population of black holes, which may exist as 
remnants of the first stars, in the Milky Way (MW) halo.  
This study would have required a dynamic range that is 
unachievable even with dissipationless numerical simulations.  
In addition, such a model can be used to study the effects of 
non-standard cosmological inputs on subhalo populations over 
a wide range of parameter space.  
\citet*[][ \citeauthor{zentner03} hereafter]{zentner03} used such 
an approach to estimate the influence of modifications to the 
primordial power spectrum on the populations of satellite halos 
and the prospects for using satellite halo populations 
to constrain cosmology.

More closely tied to the specific aims of the papers in this series,
analytic substructure models can provide an important ingredient for
modeling the large-scale clustering of galaxies by predicting halo
occupation distributions.  Coupled with a relatively low-resolution,
large-volume simulation, such a model can be used to calculate galaxy
clustering statistics, including environmental dependences, in a way
that explicitly accounts for complicated non-linear effects such as
the bias of host halos with respect to the dark matter
\citep{mo_white96,jing98,sheth_etal99,seljak_warren04}.
Alternatively, a subhalo model can be utilized in conjunction with an 
analytic halo model 
\citep[see, e.g.][]{seljak00, scoccimarro_etal01} 
to make analytic predictions for clustering
statistics rapidly, for a wide range of the space of cosmological
parameters and for various assumptions about galaxy formation.  We
demonstrate this utilization explicitly in \S~\ref{sub:app}.
Importantly, when used in these ways, our subhalo model provides a
framework for rapidly testing hypotheses about the physics of galaxy
formation against the observed clustering properties of galaxies.  Of
equal importance, the results of the semi-analytic model predictions
are easy to dissect and understand in a physical way.  We explore
these avenues in the subsequent papers of this series.

The outline of this paper is as follows.
In \S~\ref{sec:simulations}, we briefly describe three sets of
$N$-body simulations that we use to test our analytic 
subhalo model over a broad range of host halo properties.
In \S~\ref{sec:model}, we describe all of the ingredients of the
substructure model.  The model is an extension of the work of
\citeauthor{zentner03} and we pay particular attention to differences
and improvements upon that work.  As we discuss below, our 
model is similar in many respects to the model 
developed by \citet{taylor_babul01, taylor_babul04}, and 
shares common ingredients with the models of 
\citet{benson_etal02}, \citet{oguri_lee04}, 
and \citet*{vdb_etal04}.  We present our results and compare
them to the results of high-resolution numerical simulations in
\S~\ref{sec:results}.  In this section, we also discuss our
results in the context of many recent observational 
and theoretical developments and more generally describe 
the utility of our substructure model.  
We draw conclusions in \S~\ref{sec:conclusions}.
In a forthcoming companion paper (A. A. Berlind et al., in
preparation), we use this model to address the origin of the 
nearly power-law galaxy correlation function and 
several other features of galaxy clustering statistics.

Throughout this paper we refer to self-bound subunits that are 
within the virial radius of another, larger halo as 
{\em subhalos} or {\em satellites}.  
We refer to halos that are not contained within
another, larger halo, and therefore are not classified as 
subhalos, as {\em host} halos.  
We use the term {\em halo} to refer to both host
halos and satellite halos.

%
%
%
\section{Numerical Simulations} \label{sec:simulations}

The logic behind our use of numerical simulations is as follows.  
By their nature, semi-analytic models attempt to model 
complicated, non-linear dynamical processes using simple 
prescriptions and rely on inputs distilled from 
numerical simulations.  Some of the inputs for the evolution 
of subhalos is provided by controlled (non-cosmological) 
simulations \citep[e.g.,][]{colpi_etal99, velazquez_white99, 
taylor_babul01, mayer_etal02, hashimoto_etal03, taffoni_etal04, 
kazantzidis_etal04b}.  
However, cosmological simulations are still needed to specify 
the initial orbital parameters of accreted subhalos and the 
structure of cosmological halos.  We assume halo density 
profiles derived from cosmological simulations for a wide 
range of masses and cosmologies as described below.  We 
determine the initial orbital parameters of subhalos from 
a high-resolution simulation of three nearly MW-size halos.  
We subsequently compare the results of our model with 
the subhalos in simulations of a very high resolution 
cluster-size halo and a large number of halos in a 
cosmological simulation with uniform mass resolution and 
a different power spectrum normalization.  These comparisons 
over a wide range of halo masses and different power 
spectrum normalizations are thus non-trivial tests of 
the applicability of our model.

In both our simulations and our semi-analytic modeling, 
we quantify the size of host dark matter halos by their virial 
masses $\Mvir$ or, equivalently, their virial radii $\Rvir$.  
We define the halo virial radius as the radius within which 
the mean density is equal to the virial overdensity $\Dvir$, 
multiplied by the mean matter density of the universe $\rhomean$, 
so that $\Mvir = 4 \pi \rhomean \Dvir \Rvir^3 /3$.  
For the simulated halos, this radius is 
centered on the particle with the highest local 
density.  The quantity $\Dvir$ can be estimated from the 
spherical top-hat collapse approximation \citep{eke_etal98}.  
We compute $\Dvir$ using the fitting function provided by 
\citet{bryan_norman98}.  In the $\Lambda$CDM cosmology that 
we consider, $\Dvir(z=0) \simeq 337$ and tends toward the 
standard CDM (i.e. $\Omegam=1$) value $\Dvir \rightarrow 178$ 
at high redshift ($z \gsim 1$).  

All of our simulations were performed with the
Adaptive Refinement Tree (ART) $N$-body code
\citep[see][]{kravtsov_etal97,kravtsov99}. 
The simulated Galaxy-size halos 
are the three halos discussed in detail in 
\citet*[][hereafter \citeauthor{klypin_etal01}]{klypin_etal01} 
and \citet*[][hereafter \citeauthor{kravtsov04b}]{kravtsov04b}.  
Briefly, \citeauthor{klypin_etal01} used the ART code to model the 
evolution of three MW-size halos in a standard 
``concordance,'' flat, $\Lambda$CDM cosmology with 
$\Omegam = 1-\Omegal = 0.3$, $h=0.7$, and $\sigma_{8} = 0.9$ 
in a comoving box of size $25 \hMpc$ on a side.  
The simulation began with a uniform $256^3$ grid covering 
the entire computational volume.  Higher force resolution was 
achieved about collapsing structures by recursive refinement 
of all such regions using an adaptive refinement algorithm.  
The grid cells were refined if the particle mass contained 
within them exceeded a certain specified threshold value.  

The multiple mass resolution technique was used to set up the initial
conditions.  A Lagrangian region corresponding to two virial radii
about each halo was re-sampled with the highest resolution particles
of mass $m_{\mathrm{p}} = 1.2 \times 10^{6} \hMsun$, corresponding to
$1024^3$ particles in the box, at the initial timestep, $z_i=50$.  The
high mass resolution region was surrounded by layers of particles of
increasing mass with a total of five particle species.  Only the
regions containing the highest-resolution particles were adaptively
refined.  The maximum level of refinement in the simulations
corresponded to a peak formal spatial resolution of approximately $100
\hpc$.  The three MW-size halos, referred to as $G_1$, $G_2$, and
$G_3$ in \citeauthor{kravtsov04b}, have virial masses 
of $1.13 \times 10^{12} \hMsun$, $1.14 \times 10^{12}
\hMsun$, and $1.45 \times 10^{12} \hMsun$ respectively, and each of
them is resolved with approximately $\sim 10^{6}$ highest-resolution
particles within their virial radii.  Further details can be found in 
\citeauthor{klypin_etal01} and \citeauthor{kravtsov04b}.  

In addition to these three MW-size halos, we compare our model 
results to the results of a high-resolution simulation 
of a cluster-size halo of virial mass 
$\sim 1.7 \times 10^{14} \hMsun$.  
The cluster was simulated in a comoving box of $80 \hMpc$ on a 
side using the same multiple mass resolution technique 
described above in the same concordance  $\Lambda$CDM cosmology, 
as described in \citet{tasitsiomi_etal04}.
 The peak formal resolution was 0.6~$\hkpc$ and 
the mass of the highest resolution particles was 
$m_{\mathrm{p}}=3.95 \times 10^{7} \hMsun$.  The cluster was 
resolved with $\approx 4.4 \times 10^6$ particles within its 
virial radius.  This high-resolution simulated cluster halo 
was referred to as CL2 HR by \citet{tasitsiomi_etal04}.  

Lastly, we compare  our model predictions  to the subhalo populations 
in a cosmological simulation with uniform mass resolution 
containing a large number  of host  halos over  a wide  range masses.   
This simulation followed the
evolution  of  $512^3$   particles in an   $80   \hMpc$  comoving box,
corresponding to a particle mass of $m_{\mathrm{p}} = 3.16 \times 10^8
\hMsun$.  The cosmological  model was the  same $\Lambda$CDM cosmology
as  above,   but  with a  ``low''    power  spectrum normalization  of
$\sigma_8=0.75$.  The peak formal spatial resolution of the simulation
was $1.2  \hkpc$~comoving.   At $z=0$   the  simulation box   contains
$\approx 1.2 \times 10^5$ host halos with $\Mvir > 10^{11} \hMsun$ and
$\approx 3 \times 10^4$ subhalos with a bound mass $> 10^{11} \hMsun$.
We  use   this simulated halo  and subhalo   population  to  perform a
comprehensive test of our model.  This simulation was described in
\citet{kravtsov04a}, where the simulation was referred to 
as $\lcdme$.

In all  three of the numerical  simulations described above, halos and
subhalos were  identified using a variant of  the Bound Density Maxima
Algorithm \citep{klypin_etal99b}.  The  first steps of the halo-finding
algorithm  are  to compute  the local  overdensity at each particle
position using  a smoothing kernel of  $24$ particles and to find the
locations of  local maxima in  the  density field.  Starting  with the
highest  density  particle and proceeding  toward  lower density, each
density peak is marked as a potential halo  center and then surrounded
by a sphere of fixed radius $r_{\mathrm{find}}$.  All particles within
$r_{\mathrm{find}}$ are   excluded  from  further  consideration    as
potential   halo     centers.    The   search      radius    parameter
$r_{\mathrm{find}}$,  is set according to  the  size  of the  smallest
object that we aim to identify.   For the simulated MW-size halos this
was set to $r_{\mathrm{find}}=10 \hkpc$, while for the high-resolution
cluster   simulation    and    the   $\lcdme$    simulation  we    set
$r_{\mathrm{find}}=50    \hkpc$.  After    identifying  potential halo
centers, we analyze the  surrounding particles and iteratively  remove
particles   that   are    unbound  \citep[see][]{klypin_etal99b}.  All
remaining  bound particles  are   then  used to  compute  halo/subhalo
properties     such as  mass   $M$,    the  circular  velocity profile
$\Vcirc(r)=\sqrt{GM(<r)/r}$, and the peak circular velocity $\vmax$.

For subhalos, the outer boundary or outer radius  of the system can be
somewhat ambiguous and dependent upon a particular definition.
Our convention is to adopt a truncation  radius $\rt$ for subhalos, at
which  the logarithmic slope of the  density  profile constructed from
the bound particles becomes greater than  a critical value of 
$\dd \ln(\rho)/\dd \ln r = -0.5$.  
The iterative removal of unbound particles is imperfect, 
and for subhalos some fraction of unbound particles with 
nearly constant density is left on the outskirts of the subhalo.  
Some of these particles are from the diffuse mass of the 
background host halo and often some are unbound particles 
from the tidal debris of the subhalo itself.  Generally, 
the density profiles of the tidally-truncated subhalos have an 
outer slope steeper than $\dd \ln(\rho)/\dd \ln r = -3$ until the 
radius where the density of background particles becomes comparable 
to the density of the residual background particles not removed by 
the unbinding procedure.  At this radius, the profile flattens 
significantly and this is the radius that we identify as the subhalo 
truncation radius.  
The above criterion is thus based on the fact that we
do not  expect the density profiles of  CDM halos to be shallower than
this and, empirically,     this definition of   truncation   radius is
approximately equivalent  to the radius  at  which the density of  the
bound   particles  is  equal  to the background host halo density.  
We note that the outer profiles of subhalos generally fall off 
with $\dd \ln(\rho)/\dd \ln r < -3$, so the bound mass converges 
well before the truncation radius is reached.  Consequently, 
any uncertainty in the truncation radius does not translate into 
an appreciable error in the bound mass of the subhalo.  
However, throughout  most  of  this work,  we  quantify  the  size  of
satellite halos by their maximum  circular velocities $\vmax$, because
this quantity is measured more robustly and is not subject to the same
ambiguity as  a particular  mass  definition.  This makes  our results
easier to compare to the  results of other researchers using different
subhalo identification algorithms and mass definitions.

%
%
%

\section{A Model for Halo Substructure} \label{sec:model}

In order to  determine the  substructure  properties of a dark  matter
halo we must model its mass  accretion history as  well as the orbital
evolution of all accreted subhalos once they are incorporated into the
host  system.  We model substructure   using a semi-analytic technique
that incorporates  simplifying approximations  and empirical relations
observed in  numerical  simulations.  The  model  is   an updated  and
improved version of  the model described in \citeauthor{zentner03} and
in \citet{koushiappas_etal04} and    draws  on the  earlier   work  of
\citet*[][ \citeyear{bullock_etal01a}]{bullock_etal00}.  The  model of
\citeauthor{zentner03} and the   improved model that we  describe here
are  similar  in  many respects  to the models developed  by  
\citet[][ \citeyear{taylor_babul04}]{taylor_babul01},
\citet{benson_etal02}, \citet{oguri_lee04}, and \citet{vdb_etal04}.  
In this section, we give a brief description of the specific
model that we use, highlighting additions and improvements.  
We begin with a brief review of the generation of merger histories 
followed by a discussion of our assumptions about the density 
profiles of CDM halos and subhalos.  We conclude this section 
with a description of our initial conditions for subhalo orbits 
and our treatment of subhalo dynamics within the host potential. 

\subsection{Merger Trees} \label{sub:merger_trees}

The first step in determining the properties of halo substructure 
is to model the mass accretion history of the host dark matter 
halo.  A {\em merger tree} that approximates many of the 
results of $N$-body simulations can be constructed from the 
linear power spectrum of density perturbations using a 
statistical Monte Carlo technique based on the 
extended Press-Schechter (EPS) formalism 
\citep*[][ hereafter \citeauthor{lacey_cole93}; 
\citeauthor{lacey_cole94} 
\citeyear{lacey_cole94}]{bond91, lacey_cole93}.  
Using this method, we generate a list of the 
masses and accretion redshifts of all subhalos that 
have merged to form a host halo of a given mass at a 
given redshift.  In this way we are able to track both 
diffuse mass and subhalo accretion via 
mergers for each host.  By generating a large number of 
such merger trees, we can attempt to sample the diverse 
mass accretion histories that result in a host halo 
of a given fixed mass at the time at which we are 
interested in ``observing'' the halo (most often, $z=0$).  

Several authors have explored implementations of 
specific variations of the EPS formalism to construct 
realistic merger trees 
\citep[e.g. \citeauthor{lacey_cole93};][]
{sheth_lemson99,cole_etal00}.  
We employ the particular algorithm 
advocated by \citet{somerville_kolatt99}, which conserves 
mass explicitly.  Let $\sigma(M)$ represent the {\sl rms}
fluctuation amplitude in the density field smoothed on a 
scale $R$ such that the mean mass contained within a 
sphere of radius $R$ is $M$, that is 
$M = 4 \pi \rhomean R^{3}/3$ where 
$\rhomean$ is the mean mass density of the Universe.  
Following the notation established in \citeauthor{lacey_cole93}, 
let us denote $S(M) \equiv \sigma^2(M)$, 
$\Delta S \equiv S(M)-S(M+\Delta M)$, 
$w(t) \equiv \deltacrit(t)$, and 
$\delta w = w(t)-w(t+\Delta t)$, where 
$\deltacrit(t)$ is the linear overdensity for collapse 
at time $t$ in our choice of cosmological model 
($\deltacrit \approx 1.68$, see 
\citeauthor{lacey_cole93}; 
\citeauthor{white_96} \citeyear{white_96}).  
With these definitions, the probability that a halo 
of mass $M$ at time $t$ accretes an amount of mass 
associated with a change $\Delta S$ in a timestep 
corresponding to $\delta w$ is given by 

%
%
\beq
\label{eq:PofdM}
P(\Delta S , \delta w)\dd(\Delta S) = 
\frac{\delta w}{\sqrt{2\pi}(\Delta S)^{3/2}}
\exp{\frac{-(\delta w)^2}{2\Delta S}}
\dd({\Delta S})\textrm{.}
\eeq
%
%
Following \citet{somerville_kolatt99}, we generate merger 
histories by starting at the redshift that we wish to 
observe the final halo (usually $z=0$) and step backward in 
time.  At each timestep, we select halo progenitors 
by drawing values of $\Delta M$ from the distribution 
of Equation~(\ref{eq:PofdM}).  In this way, we
generate a list of progenitor masses and accretion redshifts 
at each time step.   If the minimum halo mass 
that we wish to track is $\Mmin$, then in 
order to reproduce the conditional mass function of 
EPS theory we must choose a timestep such that 
$\delta w = \sqrt{f_{\mathrm{ts}} \Mmin \dd S (\Mmin) / \dd M}$ 
with $f_{\mathrm{ts}} \ll 1$.  In practice, we choose 
$f_{\mathrm{ts}} = 10^{-2}$ as an adequate compromise 
between accuracy and computation time.  We retain all 
information about mergers with $\Delta M \ge \Mmin$ 
and treat events with $\Delta M < \Mmin$ as 
diffuse mass accretion.  In what follows, we employ several 
different values of $\Mmin$, depending upon the minimum 
halo size that we wish to track in each case.  

\subsection{Halo Mass Density Profiles}\label{sub:concentrations}

After accretion, each subhalo is subject to various 
dynamical processes as it orbits within 
the host halo.  The amount of mass that remains bound to the 
subhalo and whether or not the subhalo survives as a distinct, 
self-bound system depend upon the density structure of both the 
subhalo and the host halo.  In this subsection we 
describe our assumptions about the mass density profiles 
of CDM halos.

We characterize the size of host halos by their virial 
masses $\Mvir$ and virial radii $\Rvir$, 
as described in \S~\ref{sec:simulations}.  
In the interest of simplicity, we model all halos with 
the spherically-averaged density profile of 
\citet*[][ \citeauthor{navarro_etal97} hereafter]{navarro_etal97}:

%
\beq
\label{eq:nfw_profile}
\rho(r) = \rho_{\mathrm{s}} \left(\cvir \frac{r}{\Rvir}\right)^{-1}
\left(1 + \cvir \frac{r}{\Rvir} \right)^{-2} ,
\eeq
%
%
where $\cvir$ is the concentration parameter, which 
describes the scale radius $\rs \equiv \Rvir/\cvir$, where
the logarithmic slope of the profile, $d\ln\rho/d\ln r=-2$.  
The normalization is 
fixed by the requirement that the mass interior to $\Rvir$ be 
equal to $\Mvir$.  The NFW profile can be rewritten as a circular 
velocity profile $\Vcirc^2(r) = \Vvir^2 \cvir g(x)/x g(\cvir)$, 
where $x \equiv r/\rs$, the function $g(y) \equiv \ln (1+y) - y/(1+y)$, 
and $\Vvir \equiv \sqrt{ G \Mvir / \Rvir }$ is the halo virial velocity.  
The maximum of the circular velocity profile occurs at a radius 
$\rmax \simeq 2.16 \rs$, at a value of 
$\vmax^2 \simeq 0.22 \Vvir^2 \cvir/g(\cvir)$.

The concentrations of dark matter halos observed in simulations 
are tightly correlated with their mass accretion histories 
\citep*[][hereafter \citeauthor{wechsler_etal02}; 
\citeauthor{zhao_etal03} \citeyear{zhao_etal03}]{wechsler_etal02}.
We account for the correlation between the concentrations of 
host halos and their mass accretion histories using the proposal 
of \citeauthor{wechsler_etal02}.  
Specifically, we assign to each host halo a {\it collapse 
epoch} $\ac$, by fitting the mass accretion history of the halo, 
computed as described above in \S~\ref{sub:merger_trees}, 
with the functional form 

\beq
\label{eq:Mofa}
\Mvir(a) = \Mvir(\aobs) \exp (-2 \ac [\aobs/a-1]), 
\eeq
where $\aobs$ is the scale factor at which we observe the halo 
to have mass $\Mvir(\aobs)$.  From the fitted value of $\ac$, 
we assign each halo a concentration through 

\beq
\label{eq:cofac}
\cvir = 5.125 \Bigg( \frac{\aobs}{\ac} \Bigg) .
\eeq  
The mass accretion histories of the subhalos are generally more 
poorly sampled than the accretion histories of hosts 
because subhalos have significantly smaller 
masses than their hosts and are built by 
correspondingly fewer mergers above $\Mmin$.  
It is possible to sample the merger trees of halos and 
smaller subhalos roughly uniformly by lowering the value 
of $\Mmin$ for the smaller subhalos but this leads to 
a rapid increase in the computational expense of generating 
the merger trees and entails calculating mergers with 
halos that are far less massive than any halos that we expect 
to host luminous galaxies.  
Instead, we assign concentrations to subhalos in the following way.
If the subhalo mass is greater than $10^{3} \Mmin$, then we nearly
always have a well-sampled mass accretion history with $\gsim 50$
merger events and we use the procedure of \citeauthor{wechsler_etal02}
as outlined above.  Otherwise, we use the analytic model of \citet*[][
\citeauthor{bullock_etal01} hereafter]{bullock_etal01} to compute the
mean concentration for the mass of each subhalo at the time of
accretion.  We then compute the actual value of $\cvir$ that we assign
to each subhalo by selecting randomly from a log-normal probability
distribution for $\cvir$ with standard deviation $\sigma ( \log
(\cvir) ) = 0.14$.  \citeauthor{bullock_etal01} and
\citeauthor{wechsler_etal02} determined this to be a good
approximation for the statistical scatter of $\cvir$ for halos of
fixed virial mass in their simulations.

\subsection{Orbital Evolution of Subhalos}\label{sub:evolution}

With the accretion histories and the density 
profiles of the host and all accreted subhalos in place, the 
next step is to track the orbital evolution of each accreted 
system to determine how severely each satellite halo is 
affected by dynamical friction and tidal mass loss.  

The first step in tracking the subhalo evolution is to 
assign the initial parameters of the subhalo orbit.  
Upon accretion onto the host, each subhalo is assigned an 
initial orbital energy based on the range of binding energies 
observed in numerical simulations.  
We use the simulations of the MW-like halos $G_1$, $G_2$, 
and $G_3$ (\citeauthor{klypin_etal01,kravtsov04b}) 
to determine the initial orbital parameters for subhalos 
at each accretion event.  For each accretion event, we located 
the timestep in the simulation at which the most bound particle 
in the subhalo first appeared within the virial radius of the 
main host.  We then measured the orbital energy 
and angular momentum of the infalling subhalo.  Note that because
outputs of only $96$ epochs were saved for the 
simulation, each infalling subhalo does not have exactly the 
same relation to the host halo at the time that we measure 
these parameters and this may contribute some degree of 
uncertainty to the initial conditions.  
As in the previous work of 
\citet{klypin_etal99} and \citet{bullock_etal00}, we find that 
the initial energy distribution of orbits can be well-described 
by placing each satellite on an initial orbit with an energy
equal to the orbital energy of a {\em circular} orbit of radius 
$\Rcirc = \eta \Rvir$, where $\Rvir$ is the virial radius of 
the host halo at the time of accretion, and $\eta$ is drawn 
from a uniform distribution on the interval $[0.6,1.0]$.  

\begin{figure}[t]
\epsscale{1.0}
\plotone{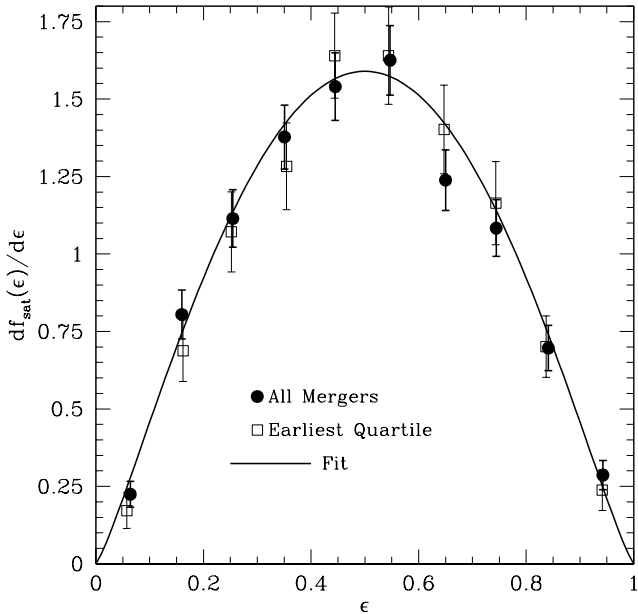}
\caption{
Input orbital circularity distribution for infalling 
subhalos measured from the high-resolution 
simulations of the $G_1$, $G_2$, and $G_3$ MW-size halos discussed 
in \citeauthor{kravtsov04b}.  
The {\it filled circles} show the measured distribution for all 
merging subhalos at the time of accretion.  The {\it open squares} 
represent the distribution measured from only the earliest $25\%$ of all 
mergers.  The {\it solid} line shows a fit of the distribution of 
all mergers to the $\beta$-distribution given by Equation (\ref{eq:beta_dist}) 
with $a=2.22$ (see main text for details).
}
\label{fig:eps}
\end{figure}

It is very important to describe the angular momenta of 
satellite orbits correctly in order to estimate the degree 
to which the subhalos are affected by tidal mass loss because 
mass loss is most rapid at pericenter, leading to a 
very strong preference for subhalos on highly eccentric 
orbits to be disrupted by tides 
\citep[for examples, see][]{taylor_babul01, taylor_babul04}.
We parameterize the initial angular momenta of subhalos 
with the orbital circularity $\epsilon$, defined 
as the angular momentum in units of the angular 
momentum of a circular orbit of the same energy, 
$\epsilon \equiv J/\Jcirc$.  The circularity parameter is 
clearly restricted to the interval $0 \le \epsilon \le 1$.  
In Figure~\ref{fig:eps}, we show the measured 
distribution of $\epsilon$ from the mergers experienced 
by $G_1$, $G_2$, and $G_3$ in terms of the fraction 
of in-falling satellites with a given initial circularity, 
$\dd \fsat(\epsilon)/\dd \epsilon$.  We find 
the distribution of orbital circularities to be 
well-fit by a one-parameter $\beta$-distribution, 

%
%
\beq
\label{eq:beta_dist}
\frac{\dd\fsat(\epsilon)}{\dd\epsilon} = 
\frac{\Gamma(2a)}{\Gamma^2(a)} \epsilon^{a-1}(1-\epsilon)^{a-1}
\textrm{,}
\eeq
which is valid on the interval $0 \le \epsilon \le 1$ and has 
the simple properties of the mean $\langle \epsilon \rangle = 1/2$, 
and standard deviation $\sigma(\epsilon) = 1/2\sqrt{2a+1}$.  Our 
best fit of $a=2.22$ is shown as the solid line in 
Fig.~\ref{fig:eps}.  The distribution of Eq.~\ref{eq:beta_dist} 
matches the measured mean 
$\langle \epsilon \rangle_{\mathrm{sim}} \simeq 0.495$ and standard 
deviation $\sigma_{\mathrm{sim}}(\epsilon) \simeq 0.219$ 
($\sigma_{\mathrm{fit}}(\epsilon) \simeq 0.214$) of subhalo 
orbital circularities in the simulation quite well.  
As we were completing our work, a study of the initial orbits of 
satellite halos was published by \citet{benson04}, the results 
of which appear to be in approximate agreement with our findings.

%
%
%
%
%
%

The distributions of initial orbits that we 
have described were culled from simulations of three 
MW-size halos and may not be generic.  
However, over the subhalo mass range that we are able to probe, 
we detect no statistically-significant variation in the 
input distributions with mass.  Additionally, 
we observe no significant evolution of these 
distributions with redshift.  An 
example of this last point is shown in Fig.~\ref{fig:eps}, 
where the open squares represent the circularity distribution 
measured for the subhalos with the earliest 25\% of 
accretion epochs.  The agreement with the full distribution is 
apparent.  We proceed by assuming that we can set initial conditions 
by drawing randomly from the energy and angular momentum 
distributions that we have described for {\em all} merging 
satellites of {\em all} masses at {\em all} redshifts.  
The comparisons with simulations discussed below show that 
this assumption appears to be reasonable. 

We evolve the orbits of satellites by treating them as 
point particles under the influence of the spherical, 
NFW potential of the host halo with an added force term 
to approximate the effect of dynamical friction.  As in 
\citeauthor{zentner03}, we account for dynamical friction by 
using a modified form of the Chandrasekhar formula 
\citep{chandrasekhar43, binney_tremaine87}.  The 
approximations of Chandrasekhar result in a 
frictional force exerted opposite the velocity of the 
satellite of magnitude 

%
%
\beq
\label{eq:chandrasekhar}
F_{\mathrm{DF}} \simeq 
\frac{4\pi G^2 \Msat^2 \rho_{\mathrm{host}}(r)}
{V_{\mathrm{sat}}^2} \ln(\Lambda) 
\Bigg( \mathrm{erf}(x)-\frac{2x}
{\sqrt{\pi}}\exp(-x^2) \Bigg)
\textrm{,}
\eeq
where $r$ is the position of the satellite halo relative 
to the center of the host potential, 
$\Msat$ is the bound mass of the satellite, 
$\rho_{\mathrm{host}}(r)$ is the mass density 
of the host halo at this position, $V_{\mathrm{sat}}$ 
is the velocity of the satellite, and 
$x \equiv V_{\mathrm{sat}}/\sqrt{2\sigma(r)^2}$, where 
$\sigma(r)^2$ is the one-dimensional velocity dispersion of 
particles in the host halo.  For the NFW host, we estimate 
the one-dimensional velocity dispersion of the host particles 
by assuming an isotropic dispersion tensor; a simple 
fitting formula for $\sigma(r)$ obtained in this way is 
given by \citeauthor{zentner03}.  We assign the Coulomb 
logarithm $\ln(\Lambda)$, according to the 
prescription of \citet*{hashimoto_etal03}, which they 
found to provide a good match to the results of their 
$N$-body experiments for individual orbits:

%
%
\beq
\label{eq:coulomb}
\ln(\Lambda) = \ln \Bigg( \frac{r}{\Rsat} \Bigg) + 
I(\Rsat/r_{\mathrm{s}})
\textrm{,}
\eeq
where $r$ is the radial position of the satellite, 
$\Rsat$ is the radius of the satellite 
(which we identify with the truncation radius as described 
below), $\rs$ is the scale radius of the NFW 
profile of the satellite halo, and the function 
$I(\Rsat/r_{\mathrm{s}})$ is an integral 
of the scattering of background particles with impact 
parameters $b < \Rsat$ 
over the radial density profile of the subhalo, 
following the method of \citet{white76}.  A convenient 
fitting  function for $I(\Rsat/r_{\mathrm{s}})$ 
for NFW profiles is given in \citeauthor{zentner03} 
[see their Eq.~(8)]. 

As in \citet{taylor_babul01} and \citeauthor{zentner03}, 
we estimate mass loss due to 
tidal stripping using a tidal approximation.  The first 
step is to estimate the tidal radius $\rtide$, as the 
radius at which the tidal force from the host 
balances the attractive force of the satellite halo 
\citep[e.g.][]{king62}.  At each timestep in the 
subhalo orbit, we estimate the mass outside of the tidal 
radius $\Msat(>\rtide)$, and strip mass on a 
timescale  proportional to the inverse of the angular 
velocity $\omega^{-1}$, of the satellite.  Specifically, 
the change in mass associated with the satellite at 
each timestep is 

%
%
\beq
\label{eq:mlr}
\delta \Msat = \alpha \Msat(>\rtide) (2\pi \delta t/\omega^{-1})
\textrm{.}
\eeq
\citeauthor{zentner03} motivated the 
$\sim 2\pi \omega^{-1}$ choice of timescale by examining 
the typical orbital energies of tidally-stripped 
material in controlled simulations \citep{johnston98}.  
The parameter $\alpha$ is intended to absorb many of 
the complicated details of the process of tidal 
mass loss, which is dependent upon the details of 
subhalo structure \citep[e.g.][]{kazantzidis_etal04a}, 
and $\alpha$ is the {\em only} parameter that 
we fix in order to normalize the model to simulation 
results.  We choose $\alpha=3.5$ as our normalization 
and discuss this choice further in \S~\ref{sub:vfuncs}.

Recent controlled $N$-body simulations \citep{hayashi_etal03} 
and the cosmological simulations of the $G_1$, $G_2$, and 
$G_3$ halos (\citeauthor{kravtsov04b}) indicate that as mass is 
lost from a satellite not only is mass from outside of the 
tidal radius removed, but there is a significant amount of 
mass redistribution within the tidal radius.  
\citet{hayashi_etal03} and \citeauthor{kravtsov04b} find that, 
on average, the maximum circular velocity of the stripped 
subhalos varies with bound mass in a remarkably simple way, 
namely $\vmax \propto \Msat^{\gamma}$, with 
$\gamma \sim 1/4-1/3$ and this scaling has been 
confirmed in the experiments of 
\citet[][S. Kazantzidis, private communication]
{kazantzidis_etal04b} with much higher resolution.
We attempt to take approximate account of this 
redistribution by enforcing the empirical scaling relation 
$\vmax \propto \Msat^{1/3}$.  
For each subhalo, tidal mass loss is most dramatic at each 
pericenter pass.  At each subsequent apocenter pass, 
we rescale the density profile of the subhalo by scaling 
$\vmax \propto \Msat^{1/3}$ and set a new scale radius 
according to the relationship between $\vmax$ and $\rs$ 
for field halos provided by the model of \citeauthor{bullock_etal01}.  
As mass is lost, we assume that the
shape of the profile maintains the NFW form
[Eq.~(\ref{eq:nfw_profile})] with a truncation radius $\Rsat$ that we
set equal to the radius that contains the remaining bound mass, rather
than the tidal radius, which does not vary monotonically.

As the final ingredients to our semi-analytic treatment of 
halo substructure, we impose criteria for declaring subhalos to 
be either completely tidally disrupted or centrally merged 
with the host halo, and therefore no longer identifiable as 
distinct substructure.  
We consider a subhalo completely tidally disrupted if, 
after any episode of tidal mass loss, the remaining bound mass 
of the satellite is less than the mass contained within 
its scale radius, $\Msat(<\rs)$.  We declare a subhalo 
centrally merged with its host halo if the apocenter of 
the subhalo orbit becomes smaller than $\rmerge = 5 \kpc$ 
due to dynamical friction.  This is a rather conservative 
central merger criterion, and in practice it is rarely 
important in our model because subhalos are always severely 
affected by tides long before their apocentric distances fall 
below $\rmerge$, so that they drop out of any mass or $\vmax$ 
cut that is designed to select subhalos that may potentially 
host luminous galaxies.  The purpose of this criterion is mostly 
one of pragmatism, to eliminate the integration of very tightly bound 
orbits which require very small time steps.

\subsection{The Subhalos of Subhalos}\label{sub:soshalos}

When each subhalo merges with a larger host, the subhalo 
itself may be the host of still smaller satellite halos.  
These {\it subhalos-of-subhalos} ({\it SOShalos}) can be an 
important contribution to the total subhalo population,
especially in the case of a very massive host halo, 
such as when a halo typical of a small group merges 
into the halo of a large 
cluster.  All of the interactions of the SOShalos cannot 
be accounted for in detail without leading to a very rapid 
increase in the computational expense of the calculation, 
mitigating the utility of the semi-analytic approach, 
the {\it raison d'etre} of which is speed and 
simplicity, and so some simplifying assumptions must be 
made.  

\citet{taylor_babul04} treat SOShalos using 
an efficient averaging scheme that is based on the 
assumption of self-similarity of the subhalo population 
of the primary host halo and the SOShalo populations of 
subhalos.  At the level of the main host, they 
determine the mean amount of mass lost by subhalos 
during their evolution and the fraction of subhalos with 
a radial position greater than the radius containing this 
amount mass $f_{\mathrm{s}}$.  At each merger event, they 
then strip a fraction $f_{\mathrm{s}}$ of the SOShalos 
associated with the incoming subhalo, 
starting with the most recently 
acquired SOShalos (i.e. last in, first out), 
and place them on correlated orbits 
within the main host.  The remaining SOShalos, 
the fraction $1-f_{\mathrm{s}}$ of SOShalos that 
were earliest acquired by the incoming subhalo, 
are then considered too tightly bound to be stripped 
from the subhalo by the main host and are added to 
the subhalo system.  

We adopt an alternative approximation scheme that is more
computationally expensive but does not rely upon the assumptions of
self-similarity, the ``last in first out'' stripping of SOShalos, or
the computation of a mean fraction of stripped SOShalos.  However, we
make several independent simplifying assumptions as follows.  We treat
SOShalos by considering their evolution through three distinct phases
punctuated by abrupt transitions.  First, before a subhalo is accreted
onto a larger host, it plays the role of a host halo and we simply
evolve the orbits of any SOShalos as described in
\S~\ref{sub:evolution}.  Accordingly, SOShalos can be significantly
dynamically evolved and can even fall below our minimum mass or
$\vmax$ thresholds prior to being acquired by the main host halo.  
Second, upon accretion of a subhalo onto a host, we
treat the SOShalos of the subhalo, if they are present and still above
the minimum halo mass of interest, by continuing to integrate their
orbits within the subhalo potential, neglecting their interactions
with the main host halo.  Third, we consider the possibility that a
SOShalo may be stripped from the subhalo and deposited in the main
host.  This represents a transition where the dynamics of the SOShalo
go from being dominated by the subhalo to being dominated by the main
host.  During the evolution of the subhalo in the main host, if the
tidal radius of the subhalo becomes smaller than the radial position
of a SOShalo, we declare the SOShalo to be stripped off of the
subhalo.  We then remove it from the subhalo, place it in the
potential of the main host, and continue integrating its orbit until
the final epoch (usually $z=0$).

When we place a stripped SOShalo within the potential of the main
host, we must assign it an orbit that is closely associated with the
orbit of the subhalo from which it was stripped.  \citet{johnston98}
found that the typical specific energy of tidally stripped particles 
is set by the change in the host halo potential on the length scale of the
orbiting satellite, $\Delta E_{\mathrm{ts}} = \rtide \dd
\Phi_{\mathrm{host}}(r)/\dd r $.  We set the stripped SOShalo on an
orbit with specific energy equal to the orbital energy of its 
{\em parent subhalo} in the {\em main host} $\pm \Delta E_{\mathrm{ts}}$, 
with the sign of the shift in orbital energy chosen randomly to 
correspond to the leading and trailing tails of the tidal debris.  

%
%
%
%
%
\vspace{0.1in}

\section{Model Results and Comparisons with Numerical Simulations}
\label{sec:results}

In this Section, we present the results of the substructure model
described in \S~\ref{sec:model} along with a comprehensive comparison
with the results of high-resolution numerical simulations over the
mass range where the two approaches can be compared robustly.

\subsection{Subhalo Velocity and Mass Functions} \label{sub:vfuncs}

\begin{figure}[t]
\epsscale{1.0}
\plotone{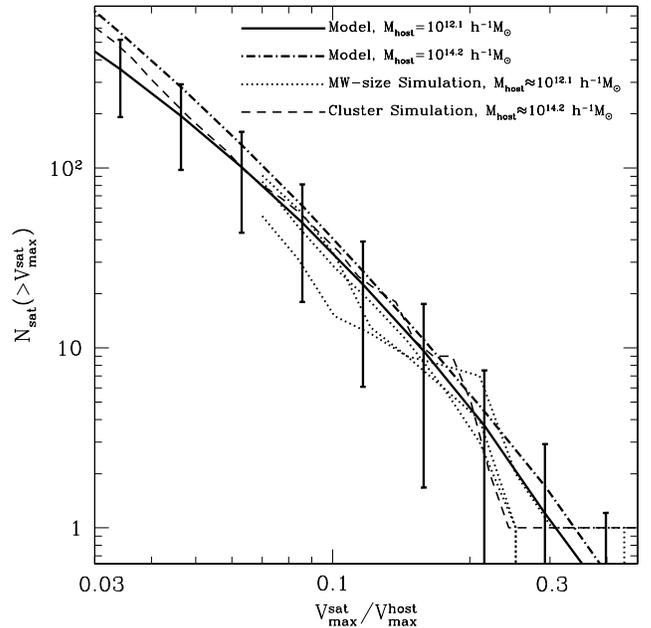}
\caption{
The mean cumulative velocity functions $\Nsg(>\vmax)$, 
of satellite halos.  The {\it solid} line depicts the 
mean value of the number of subhalos with a maximum 
circular velocity greater than $\vmaxsat$, as a function of 
$\vmaxsat$ predicted by $100$ model realizations of a 
$\Mhost = 10^{12.1} \hMsun$ host halo.  
Velocities are scaled in units of the maximum circular 
velocity of the host halo $\vmaxhost$.  The errorbars depict 
the $1-\sigma$ scatter among the $100$ model realizations.  
The {\it dotted} lines show the cumulative velocity functions 
of the three simulated MW-size halos.  The {\it dashed} line 
shows the cumulative velocity function of the simulated 
cluster-size halo \citep[CL2 HR in ][]{tasitsiomi_etal04}.  
The {\it dot-dashed} line is the mean 
predicted cumulative velocity function from $100$ model 
realizations of a $\Mhost = 10^{14.2} \hMsun$ halo.  
In the interest of clarity, the realization-to-realization 
scatter is not shown for this case but is similar to the 
scatter for the MW-size model realizations.
}
\label{fig:Vfunc}
\end{figure}

As in \citeauthor{zentner03}, the particular algorithm 
described in \S~\ref{sec:model} made use of inputs from 
numerical simulations and, in particular, was designed 
to approximately reproduce the cumulative velocity 
function (CVF) $\Nsg(>\vmax)$, of simulated MW-size halos 
(\citeauthor{kravtsov04b}).  The CVF is defined 
as the number of subhalos above a given threshold in 
maximum circular velocity $\vmax$, as a function of 
$\vmax$.  Figure~\ref{fig:Vfunc} shows model predictions 
for CVFs compared to the results of numerical simulations.  
In all cases, we refer to subhalo maximum circular 
velocities $\vmaxsat$, that are scaled by units of 
the maximum circular velocity of the host halo $\vmaxhost$, 
in order to approximately scale out the dependence of the 
velocity function on the size of the host halo and 
facilitate the comparison of velocity functions of hosts 
of various sizes.  The solid line in Fig.~\ref{fig:Vfunc} 
shows the mean CVF from $100$ model realizations of a host 
halo with virial mass 
$\Mhost = 10^{12.1} \hMsun$ in a standard 
$\Lambda$CDM cosmology with $\Omegam=1-\Omegal=0.3$, 
$h=0.7$, and $\sigma_8=0.9$.  The error bars on this 
result represent the $1-\sigma$ realization-to-realization 
scatter about the mean value measured from these 
$100$ model realizations.  The dotted lines show the 
measured cumulative velocity functions of the three 
simulated MW-size halos $G_1$, $G_2$, and $G_3$ 
introduced in \S~\ref{sec:simulations}.  The dotted 
lines are truncated at the value of $\vmaxsat$ where 
the simulation halo catalogs become incomplete due 
to finite resolution.  The good 
agreement between the model CVF and the CVFs of the 
simulated MW-size halos is apparent and was achieved 
by setting the mass loss parameter $\alpha=3.5$ 
in Eq.~\ref{eq:mlr}.  We emphasize that, as in 
\citeauthor{zentner03}, our semi-analytic model was 
normalized to {\em this statistic only} and as such, 
it is now possible to make {\em predictions} for other 
statistics, and to diagnose and assess the general 
applicability of the model by comparing these 
predictions to the results of numerical simulations.  

Figure~\ref{fig:Vfunc} also shows the first 
prediction of the semi-analytic model compared to 
the result of an $N$-body simulation.  The dot-dashed 
line shows the CVF of a $\Mhost = 10^{14.2} \hMsun$ 
cluster-size host halo in the same ``concordance''
cosmology.  For clarity, the scatter 
among these realizations is not shown, but it is similar 
to the scatter on the CVFs of the MW-size halos.  
The dashed line is the CVF of a simulated cluster-size 
halo of the same mass.  The predictions of the model and 
the simulation are in good agreement as the simulated halo 
is well within the $1-\sigma$ scatter about the mean CVF 
of the model.  Notice also that the model predicts that 
the CVF should not scale in a precisely self-similar manner 
with the size of the host.  The model results in a CVF that 
is slightly higher at higher values of $\vmaxhost$.  We 
return to this point shortly.

%
%
%
%
%
%

\begin{figure}[t]
\epsscale{1.0}
\plotone{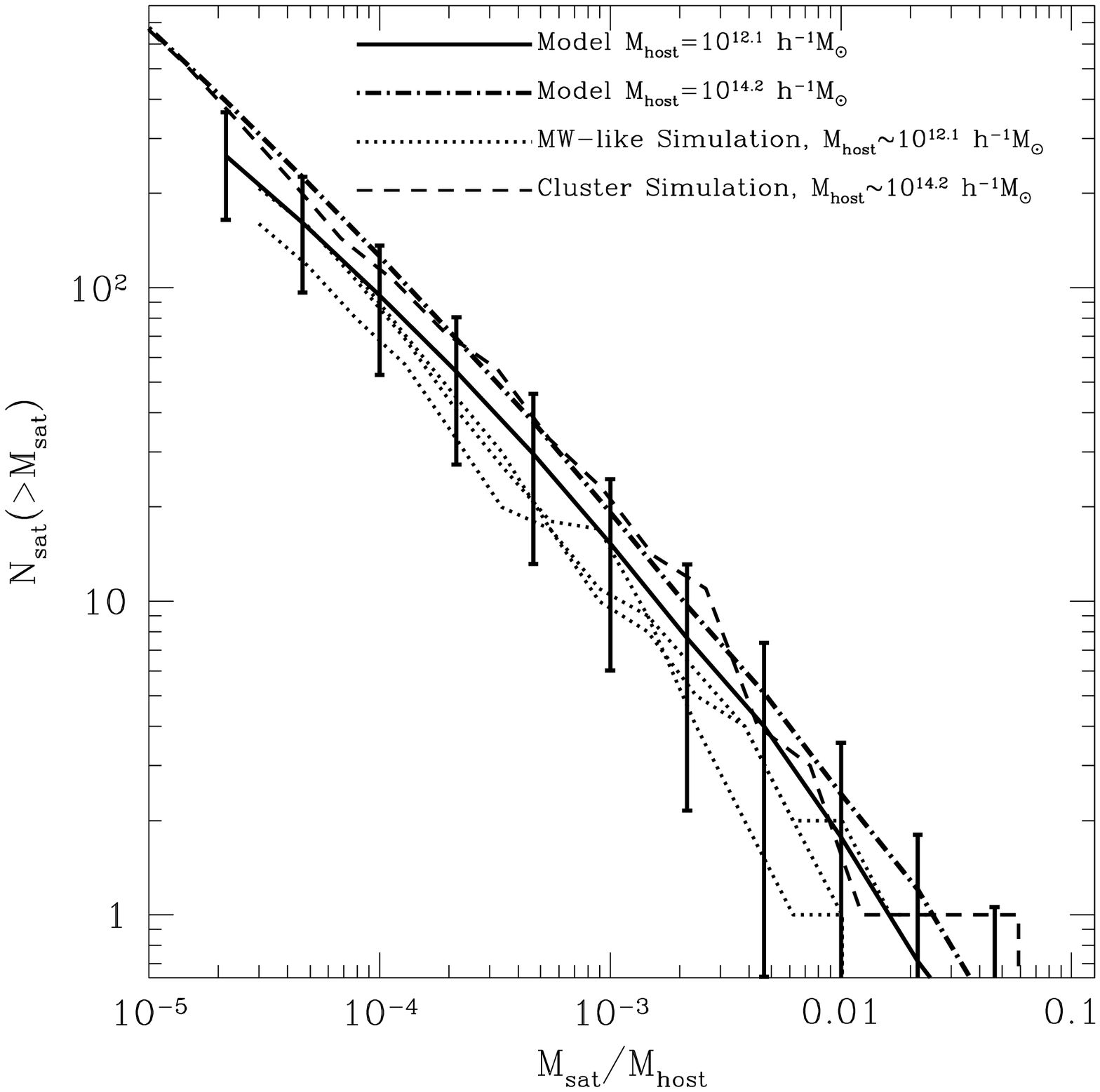}
\caption{
The mean cumulative mass function (CMF) of subhalos 
as a function of subhalo mass relative 
to the mass of the host halo, $\Msat/\Mhost$.  
The {\it solid} line shows the mean value of 
$\Nsat(>\Msat/\Mhost)$ for a $\Mhost=10^{12.1}\hMsun$ 
host computed from $100$ realizations of 
the dynamical model of \S~\ref{sec:model}.  
The errorbars depict the $1-\sigma$ scatter 
among the $100$ realizations.  The {\it dotted} lines 
show the CMFs of the three simulated halos 
of similar mass described in \citeauthor{klypin_etal01} 
and \citeauthor{kravtsov04b}.  The {\it dot-dashed} line 
shows the mean CMF of a $\Mhost=10^{14.2} \hMsun$ halo 
predicted by $100$ model realizations (for clarity, 
the scatter is not shown in this case but is similar 
to that of the smaller host).  The {\it dashed} line 
shows the CMF measured from a high-resolution 
simulation of a cluster-size halo of a similar mass
\citep[halo CL2 HR in][]{tasitsiomi_etal04}.  
}
\label{fig:Mfunc}
\end{figure}

As a further test of our model, we now turn to the 
cumulative mass function (CMF) $\Nsg(>\Msat)$ of subhalos.  
As with the CVF, we refer to subhalo masses $\Msat$ in 
units of the host halo mass $\Mhost$ in order to scale 
out the gross dependence of $\Nsg(>\Msat)$ upon the 
mass of the host halo.  We show CMFs in Figure~\ref{fig:Mfunc}.  
Again, the agreement between the results of the model and the
numerical simulations is good, lending confidence in the ability
of the model to account approximately for the dominant physical
effects that determine subhalo properties, and to be applied more
generally.  Note also that, just as we saw for the CVF in
Fig.~\ref{fig:Vfunc}, the CMFs in Fig.~\ref{fig:Mfunc} show that the
number of subhalos of a given size does not scale self-similarly
with the size of the host.  Larger host halos tend to have more
substructure at a fixed value of $\Msat/\Mhost$ (or
$\vmaxsat/\vmaxhost$), in agreement with the recent 
numerical results of \citet{gao_etal04a} and the 
simplified analytic considerations of \citet{vdb_etal04}.

%
%
%
%
%
%


\subsection{The Radial Distribution of Subhalos} \label{sub:subpro}

The radial distribution of subhalos within their hosts 
has received a significant amount of attention recently 
and is a necessary ingredient of models of galaxy 
clustering statistics \citep{seljak00, sheth_etal01, berlind02} 
and in the modeling of the flux-ratio 
anomalies in strong lens systems 
\citep*[e.g.,][]{dalal_kochanek01, bradac_etal02, chen_etal03}.  
At least three groups have recently performed 
convergence studies that demonstrate that the 
radial distribution of subhalos of all sizes is 
significantly less centrally-concentrated than 
the distribution of dark matter 
\citep{diemand_etal04, gao_etal04a, nagai_kravtsov04}, 
confirming several earlier results 
\citep[\citeauthor{ghigna_etal98} \citeyear{ghigna_etal98}, 
\citeyear{ghigna_etal00}; ][]{colin_etal99, springel_etal01}.  
\citet{diemand_etal04} and \citet{gao_etal04b} stressed that 
the observed spatial distribution of galaxies in clusters 
\citep*{carlberg_etal97, vandermarel_etal00, lin_etal04} 
appears to be more centrally-concentrated and not as 
strongly anti-biased with respect to the dark matter as 
the distribution of subhalos in dissipationless 
$N$-body simulations \citep[see also][]{springel_etal01}.  
\citet{vdb_etal04a} analyzed satellite galaxies in the 
Two-Degree Field Galaxy Redshift Survey \citep{colless_etal01} 
finding preliminary indications that observed galaxies 
do follow a centrally-concentrated, NFW-like radial distribution 
over a wide range of host halo masses, though they cannot 
rule out shallower distributions because of incompleteness 
of close pairs \citep{cole_etal01} and more detailed studies 
are required to confirm this.

\citet*{taylor_etal03} and 
\citet{taylor_babul04b,taylor_babul04c} 
used semi-analytic methods 
to argue that the distribution of subhalos should 
be more centrally-concentrated than that observed 
in simulations and that simulations may still be 
subject to the problem of ``overmerging'' in the 
central regions of halos \citep[see ][]{klypin_etal99b}.  
The results of the recent studies 
by \citet{diemand_etal04}, \citet{gao_etal04a}, and 
\citet{nagai_kravtsov04} cast doubt upon the proposal of 
\citet{taylor_etal03}.  \citet{nagai_kravtsov04} suggest
that the radial distribution of galaxies selected on luminosity
cannot be compared directly with that of subhalos selected
by mass because galaxies do not likely lose luminosity at the 
same rate that they lose dark matter while they orbit inside 
a larger host halo.  
Nevertheless, it is important to pursue further tests 
of this issue.
The growing importance of the radial distributions 
of subhalos and various subsets of subhalos with 
specific properties, 
coupled with these recent theoretical studies, 
provide interesting motivations for comparing the spatial 
distributions of subhalos in our model with that 
observed in numerical simulations.

\begin{figure*}[t]
\epsscale{1.9}
\plotone{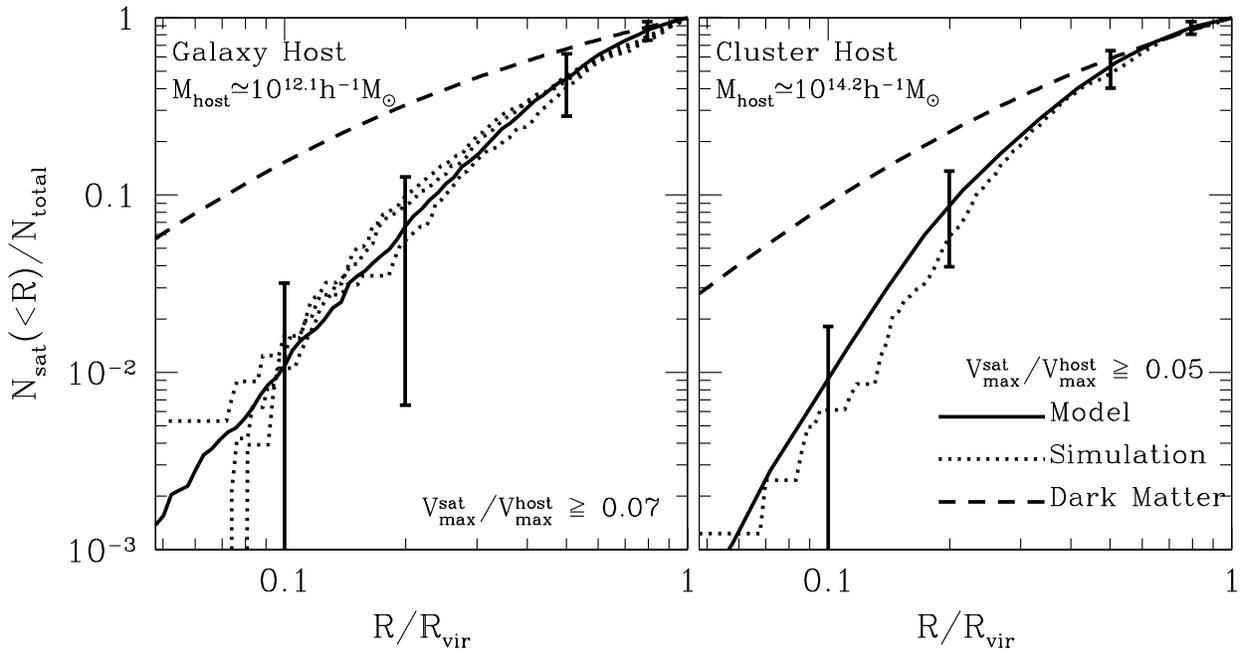}
\caption{
{\it Left Panel:} The cumulative radial distribution of 
satellite halos in MW-size host halos.  We plot the 
fraction of subhalos within a given distance from the 
host halo center $\Nsg(<R)/N_{\mathrm{total}}$ as a 
function of distance $R$, in units of the halo virial 
radius, $R/\Rvir$.  
The {\it solid} line shows the mean cumulative 
radial distribution of subhalos with 
$\vmaxsat \ge 0.07\vmaxhost$ for 
$\Mhost=10^{12.1} \hMsun$ computed from $100$ 
realizations of the semi-analytic substructure 
model described in \S~\ref{sec:model}.  
The error bars show the $1\sigma$ scatter in 
the $100$ model realizations at only four radii 
for clarity.  The {\it dotted} lines show the 
cumulative radial subhalo distributions for the 
$G_1$, $G_2$, and $G_3$ simulated MW-size halos.  
The {\it dashed} line represents the distribution 
of dark matter for an NFW halo with 
concentration parameter $\cvir=12.6$, typical of a 
MW-size halo (\citeauthor{bullock_etal01}).  The simulated 
halos have best-fitting NFW concentrations of 
$\cvir \simeq 9.5$, $12.5$, and $14.5$.  
{\it Right Panel:} The cumulative radial distribution 
of satellite halos in cluster-size host halos.  The 
line types are the same as in the {\it right panel}.  
In this case, the host mass is $\Mhost=10^{14.2}\hMsun$ 
and all subhalos with $\vmaxsat \ge 0.05 \vmaxhost$ 
are plotted.  
The {\it dashed} line shows the distribution 
of dark matter for an NFW profile with concentration 
$\cvir=6.1$, which is typical for this host mass.  
The best-fitting concentration for the simulated halo 
is $\cvir \simeq 8.5$.  In both panels, the simulated 
radial distributions were computed by stacking the 
last five simulation outputs for each object.
}
\label{fig:NltR}
\end{figure*}

In Figure~\ref{fig:NltR}, we compare the results of 
our model predictions for the radial distributions of 
subhalos with the results of the $G_1$, $G_2$, and $G_3$ 
MW-size halo simulations and the high-resolution cluster 
simulation discussed in \citet{tasitsiomi_etal04}.  
In Figure~\ref{fig:NltR}, we plot the cumulative number 
of subhalos at a distance less than $R$ away from the 
center of the host halo $\Nsg(<R)$, as a function of 
$R$ and normalize by the total number of subhalos within 
the virial radius of the halo $\Ntotal$ 
(Fig.~\ref{fig:Vfunc} and Fig.~\ref{fig:Mfunc} already 
show that the values of $\Ntotal$ are in agreement).  
In the left hand panel, we compare the radial distribution 
of subhalos within MW-size host halos.  
We choose the selection criterion $\vmaxsat/\vmaxhost \ge 0.07$ 
in order to restrict our results to subhalos that are 
sufficiently large as to be well-resolved in the simulation 
and fairly insensitive to numerical effects.  
The dot-dashed line shows the 
distribution of dark matter for an NFW profile with a 
concentration parameter of $\cvir = 12.6$.  This is a 
typical value of $\cvir$ for a MW-size halo in this 
cosmology (\citeauthor{bullock_etal01}) and  
is the mean value of the concentration of the $100$ hosts 
in the semi-analytic model.  The three simulated MW-size 
halos $G_1$, $G_2$, and $G_3$ have best-fit
concentration parameters of $\cvir \simeq 9.5$, $12.5$, 
and $14.5$ respectively.  
Similarly, in the right hand panel of Fig.~\ref{fig:NltR}, 
we compare the radial distribution of subhalos 
of the simulated cluster halo with the distribution of 
subhalos in $100$ model realizations of a 
$\Mhost = 10^{14.2} \hMsun$ host halo.  
The maximum circular velocity threshold in this case is 
$\vmaxsat/\vmaxhost > 0.05$.  
Once again, the dot-dashed line represents the 
distribution of dark matter for an NFW profile with 
$\cvir = 6.1$, which is typical for a halo of this mass and 
is close to the mean concentration of the $100$ hosts in the 
semi-analytic model.  The simulated cluster halo has a best-fitting 
concentration of $\cvir = 8.5$, which is roughly 
$\sim 1-\sigma$ higher than the mean concentration at this 
mass according to the halo-to-halo scatter 
observed in simulations (\citeauthor{bullock_etal01}).  
For the simulated subhalo distributions, we computed 
profiles by stacking the last five simulation outputs 
($a=1.00,0.99,0.98,0.97,0.96$ for $G_1$, $G_2$, and $G_3$ 
and $a=1.000,0.995,0.990,0.985,0.980$ for the cluster) 
in order to overcome the noise in the measurement.  

%
%
%
%
%
%
%
%

Several interesting features are evident in 
Figure~\ref{fig:NltR}.  
First, the anti-bias of the subhalo population 
with respect to the dark matter is evident in 
both panels of Fig.~\ref{fig:NltR} for both the 
simulated halos and the model halos.  
A MW-size halo typically has $\sim 35\%$ of its 
mass within $\sim 0.2 \Rvir$, yet it has only 
$\sim 5\%$ of its satellite halos 
located this close to its center.  
Second, notice the excellent agreement between 
the simulation and model subhalo profiles in both 
panels.  Again, this agreement is a non-trivial confirmation 
of the success and more general applicability of 
the semi-analytic model.  After designing the model 
to reproduce the {\em number} of subhalos in a 
MW-size host, the model naturally recovers 
the correct radial distribution of satellites within 
both a MW-size host halo and a more massive 
cluster-size halo.  

The agreement between the model subhalo 
distribution and the distribution of subhalos in 
the simulations is an encouraging sign that the 
semi-analytic model captures the dominant 
physical processes that determine the demographics 
of CDM subhalos.  Moreover, this indicates that the model 
of \S~\ref{sec:model} results in a spatial distribution 
of subhalos that is in conflict with the model of 
\citet{taylor_etal03} and \citet{taylor_babul04b},
 who find a much more centrally-concentrated 
distribution of subhalos.  
The radial distributions of subhalos in our model 
support the conclusion that the 
anti-bias of subhalos with respect to dark matter 
is a physical effect rather than the result of 
numerical overmerging, yet it will likely require 
further study to provide a definitive answer to this 
question. However, our results indicate that the
disagreement between semi-analytic models and 
simulation results may depend upon the specifics of 
the model implementation and is not a generic 
feature of such models. 

%
%
%
%

\subsection{The Halo Occupation Distribution of Subhalos}
\label{sub:subhod}

We now turn to a more comprehensive comparison of our semi-analytic
model to the predictions of cosmological $N$-body simulations.  To do
this, we compare the results of our model to the subhalo populations
in the $\lcdme$ simulation studied by \citet{kravtsov04a}.  Recall
that the $\lcdme$ simulation was performed for a $\Lambda$CDM
cosmological model with $\sigma_8=0.75$.  This power spectrum
normalization is lower than the normalization of $\sigma_8=0.9$ used
in the simulations of the $G_1$, $G_2$, and $G_3$ halos as well as the
CL2 cluster halo to which we compared our model in \S~\ref{sub:vfuncs} 
and \S~\ref{sub:subpro}.  We therefore run our model again using 
$\sigma_8=0.75$.  The parts of the model that are directly affected by 
this change are the EPS merger trees of halos and their mass density
profiles.  Comparing the $\lcdme$ simulation results with the model 
at this different power spectrum normalization and at different halo 
masses than the $\sigma_8=0.9$, $\Mhost = 10^{12.1} \hMsun$ simulations
that we used to tune the model is not a trivial exercise and tests the 
general applicability of the model.

\begin{figure}[t]
\epsscale{1.0}
\plotone{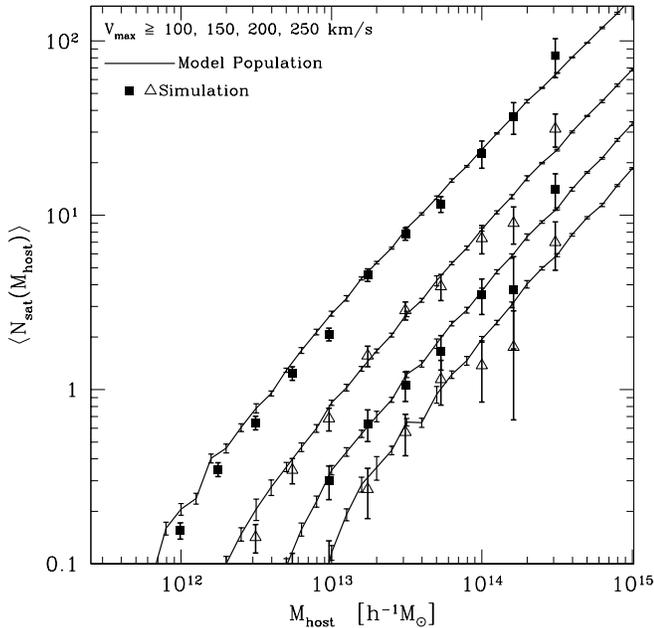}
\caption{
The mean subhalo occupation number as a function of host halo mass,
for several peak circular velocity thresholds: 
$\vmax \ge 100$, $150$, $200$, and $250 \kms$. {\it Squares} and 
{\it triangles} show results of the $\lcdme$ $N$-body simulation, 
while the results of our semi-analytic model described in \S~\ref{sec:model} 
is shown by the {\it solid curves}.  
The model results are based on $1000$ Monte Carlo 
realizations in each host mass bin, with host mass bins 
spaced by $\Delta(\log\Mhost)=0.1$.  For both the simulation 
and the model results, the error bars represent the 
estimated error on the mean value of $\Nsg$ in each 
mass bin.  
} 
\label{fig:Ncomp}
\end{figure}

Our first comparison is shown in Figure~\ref{fig:Ncomp}.  
In this plot, we show the mean number of subhalos, $\Nsat$, 
as a function of host halo mass, 
for four different $\vmax$ thresholds:  
$\vmax \ge 100$, $150$, $200$, and $250\ \kms$.  
The lines show the result of $1000$ model realizations 
in each host mass bin from $\log(\Mhost/\hMsun)=11.0$ 
to $\log(\Mhost/\hMsun)=15.0$ with bins evenly-spaced 
by $\Delta(\log \Mhost)=0.1$.  The squares and 
triangles represent the $\Nsat$ measured from the 
halos in the $\lcdme$ simulation.  

Figure~\ref{fig:Ncomp} shows a rather remarkable result: our model of
subhalo populations matches the results of the simulations very 
well at each host mass and each $\vmax$ threshold.  At a single, fixed
value of $\Mhost$, the four model or simulation points associated with
that value correspond to four points on the CVF for host halos of that
mass.  As such, Figure~\ref{fig:Ncomp} demonstrates that, after
culling the input distributions for impact energies and impact
parameters of subhalos at accretion from a cosmological $N$-body
simulation of the formation of three MW-size halos, and fixing our
one-parameter model to match the amplitude of the CVF of subhalos of a
MW-size host, the model then successfully reproduces the CVFs of
subhalos over a range of roughly three orders of magnitude in host
mass in a cosmology with a different power spectrum normalization.

%
%
%
%
%
%

\begin{figure}[t]
\epsscale{1.0}
\plotone{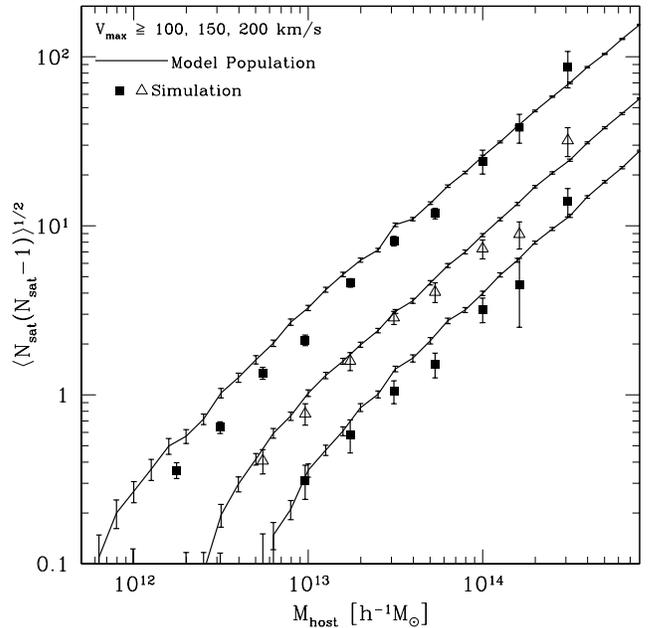}
\caption{
The square root of the second 
moment of the halo occupation distribution of subhalos 
$\langle \Nsg (\Nsg-1) \rangle^{1/2}$, as a function of 
$\Mhost$.  The {\it solid} lines are the mean of $1000$ 
realizations of the semi-analytic model of 
\S~\ref{sec:model}.  The {\it squares} and 
{\it triangles} show the results for the 
subhalo populations measured in the $\lcdme$ 
cosmological $N$-body simulation.  
From top to bottom the three sets of {\it lines} and 
{\it points} are for three different subhalo $\vmax$ 
thresholds: 
$\vmax \ge 100$, $150$, and $200 \kms$ respectively.  
}
\label{fig:NNm1}
\end{figure}

As a further diagnostic, we can go beyond the first moment, or mean,
of the occupation distribution of subhalos and explore the shape of
the probability distribution for hosting $\Nsg$ satellites at a 
given host mass $P(\Nsg \vert \Mhost)$, by comparing the second
moment of the distribution.  The higher order moments of the subhalo
distribution are important for studies of the $N$-point statistics of
subhalo (or galaxy) populations where $N \ge 2$.  In
Figure~\ref{fig:NNm1}, we show results for the second moment of the
subhalo occupation distribution $\NNsat$, as a function of host mass.
As in Fig.~\ref{fig:Ncomp}, the lines represent the model results, the
points represent the simulation results and the error bars represent
the error on $\NNsat$.  We show subhalo populations above three cuts
in $\vmax$, subhalos with $\vmax \ge 100$, $150$, $200\ \kms$.  The
agreement between the model and the simulations is good for the second
moment of the distribution of subhalo populations for the highest two
$\vmax$ cuts.  However, a notable feature of Fig.~\ref{fig:NNm1} is
the discrepancy between the model results and the simulation results
at low values of $\Mhost$ or, equivalently low values of $\NNsat$, in
the lowest $\vmax$ sample, where the model overpredicts the second
moment of the distribution by as much as $\sim 40\%$.

%
%
%
%
%
%
%
%
%
%
%
\begin{figure*}[t]
\epsscale{1.75}
\plotone{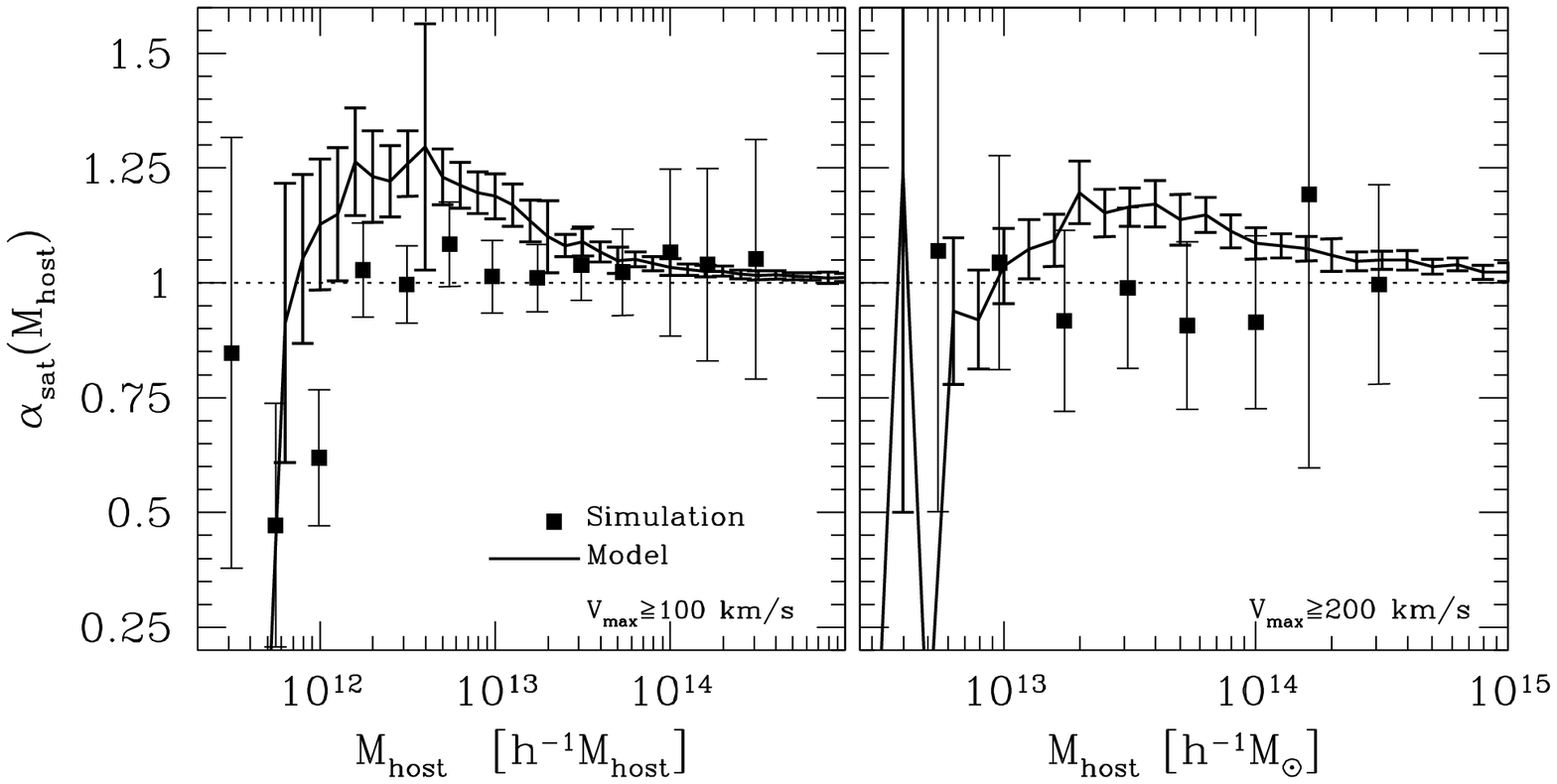}
\caption{
The width of the halo occupation distribution of 
subhalos compared to the width of the Poisson 
distribution. We plot the quantity
$\alpha \equiv \NNsat^{1/2}/\Nsat$ as a function of 
host mass for two circular velocity thresholds: $\vmax \ge 100 \kms$ 
({\it left panel}) and $\vmax \ge 200 \kms$ ({\it right panel}).  
The {\it solid} lines correspond to the results of the 
analytic model of \S~\ref{sec:model} and the {\it squares} 
correspond to simulation results.  The horizontal 
{\it dotted} line marks the value $\alpha = 1$ which 
corresponds to the Poisson distribution.
}
\label{fig:alpha}
\end{figure*}
%
%
%

\citet{kravtsov04a} found the second moment of the 
occupation distribution of subhalos to be consistent 
with that of a {\em Poisson} distribution in the 
host mass regime where $\Nsat \gsim 0.1$.  
In this case, the probability 
distribution function for a host halo of mass $\Mhost$ 
to contain a particular number of satellite halos is 
completely specified by the mean of the distribution 
because the higher order moments are related to the 
mean through 
$\langle N(N-1)\ldots(N-i) \rangle = \langle N \rangle^{i+1}$.  
In Figure~\ref{fig:alpha}, we compare the 
occupation distribution of subhalos 
predicted by our semi-analytic 
substructure model to a Poisson distribution 
directly by plotting the quantity  

%
%
\beq
\label{eq:alpha_definition}
\alpha \equiv 
\langle N_{\mathrm{sat}}(N_{\mathrm{sat}}-1) \rangle^{1/2} / 
\langle N_{\mathrm{sat}} \rangle \textrm{.}
\eeq
%
%
For a Poisson distribution, $\NNsat^{1/2} = \Nsat$ so that $\alpha =
1$, while for distributions that are narrower ({\it sub-Poisson}) or
broader ({\it super-Poisson}) than a Poisson distribution, $\alpha <
1$ and $\alpha > 1$ respectively.  Figure~\ref{fig:alpha} shows
$\alpha$ as a function of $\Mhost$ for subhalos in the semi-analytic
model and the $\lcdme$ simulation for subhalos with $\vmax \ge 100
\kms$, and $\vmax \ge 200 \kms$ in the left and right panels
respectively.  Figure~\ref{fig:alpha} shows that for the $\vmax \ge
100 \kms$ threshold, the semi-analytic model produces a probability
distribution for the number of satellites at a given host mass that is
significantly super-Poisson in the host mass range $10^{12} \lsim
\Mhost/\hMsun \lsim 10^{13}$, corresponding to a mean number of
satellites $0.1 \lsim \Nsat \lsim 1$.  In this range, the model
over-predicts $\alpha$ for the occupation distribution of subhalos by
as much as $\sim 25\%$ with respect to the simulated occupation
distribution of subhalos.  The simulated occupation distribution of
subhalos is consistent with the Poisson value of $\alpha = 1$ for
$\Mhost \gsim 10^{12} \hMsun$ ($\Nsat \gsim 0.1$).  At higher $\vmax$
thresholds, the magnitude of the discrepancy decreases, as can be seen
in the right panel of Figure~\ref{fig:alpha}, but qualitatively, we
find that the super-Poisson ``bump'' at $\Nsat \sim 0.1-1$ persists at
all $\vmax$ thresholds.  The over-prediction of the second moment of
the distribution relative to the simulation result appears to be a
fairly general feature of our semi-analytic model.  The model of 
\citet{vdb_etal04} yields a similar overestimate of the second 
moment despite the fact that their mass loss model is quite 
different from the model that we present, and is 
based on a simplified averaging of 
subhalo mass loss (F. C. van den Bosch, private communication).  
Coupling this with the fact that EPS formation times seem to 
exhibit a larger scatter than the formation times of simulated 
halos, as pointed out by \citeauthor{wechsler_etal02}, 
leads us to speculate that the 
origin of this discrepancy is a fundamental shortcoming of the 
standard EPS formalism for generating halo merger trees.  Further 
investigation is necessary to test this hypothesis.

%
%
%
%
\begin{figure}[!t]
\epsscale{1.0}
\plotone{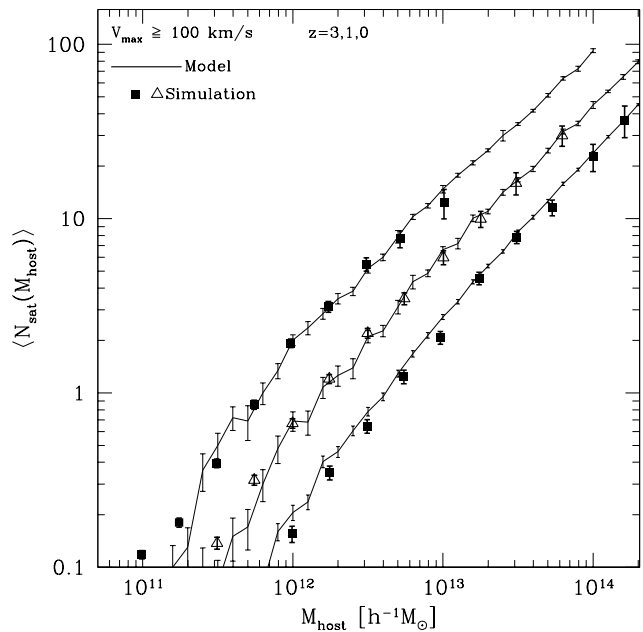}
\caption{ The redshift dependence of the mean of subhalo occupation.
  The lines show the predictions of the semi-analytic model at
  redshifts $z=0$ ({\it bottom}), $z=1$ ({\it middle}), and $z=3$
  ({\it top}).  The model results are based on $1000$ Monte Carlo
  realizations at each host mass bin at $z=0$, and $100$ realizations
  at redshifts $z=1,3$.  The {\it filled squares} and {\it open
    triangles} represent the same quantity measured from the
  simulations described in \citet{kravtsov04a}.  For both the model
  predictions and the predictions of the simulations, the errorbars
  represent the estimated error on $\Nsat$.  Note that we plot the
  subhalo occupation at only one threshold ($\vmax \ge 100 \kms$)
  shown in Fig.~\ref{fig:Ncomp} because large halos become
  increasingly rare with increasing redshift causing the statistics to
  be poor at high redshifts and large thresholds.  
}
\label{fig:Ncomphiz}
\end{figure}
%
%
%
%
%
%
\begin{figure}[!t]
\epsscale{1.0}
\plotone{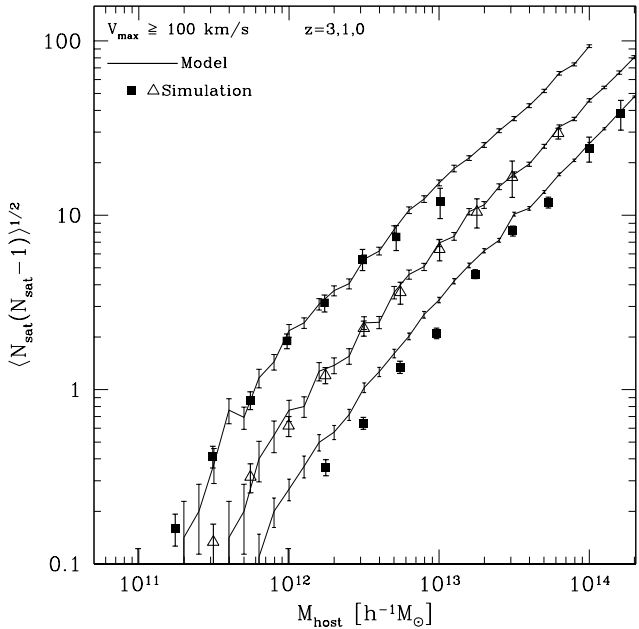}
\caption{
The redshift dependence of the second moment of the 
subhalo occupation distribution, 
$\langle \Nsg(\Nsg-1)\rangle^{1/2}$.  The lines show 
the results of the semi-analytic substructure model 
for subhalos with $\vmax \ge 100 \kms$ at redshifts 
$z=0$ ({\it bottom}), $z=1$ ({\it middle}), and 
$z=3$ ({\it top}).  The model results are based on 
$1000$ model realizations at each host mass bin at 
$z=0$, and $100$ realizations at redshifts $z=1,3$.  
The {\it filled squares} and {\it open triangles} 
represent the same quantity measured from the 
$\lcdme$ simulation.  
}
\label{fig:NNm1hiz}
\end{figure}
%
%
%

As a further test of the model, we generalize this comparison and
examine the subhalo occupancy predicted by the model as a function of
redshift.  First, Figure~\ref{fig:Ncomphiz} shows the mean number of
subhalos with $\vmax \ge 100 \kms$ as a function of host mass at
redshifts $z=0,\ 1,$~and~$3$.  We illustrate the redshift dependence
using only this one $\vmax$ threshold at high redshift because the
number of large halos decreases rapidly with redshift, causing the
statistics to be rather poor at larger thresholds.  At $z=0$ the model
points represent the same $1000$ realizations per mass bin as shown in
Fig.~\ref{fig:Ncomp}.  At redshifts $z=1$ and $z=3$, the model results
are based on $100$ realizations per mass bin.  Again, the agreement
between the model and the $\lcdme$ simulation is rather impressive.
The simple model of halo substructure that we developed in
\S~\ref{sec:model}, predicts a mean subhalo occupation number at high
redshift that is in good agreement with the predictions of the full
numerical simulation.  This result also demonstrates the clear trend
for CDM halos to host a larger amount of substructure at higher
redshifts.  Here we select halos at a fixed $\vmax$ cut; using a 
sample selected at a fixed number density, yields the same trend, 
though slightly less pronounced.

The high-redshift agreement is not limited to the mean occupation.
Figure~\ref{fig:NNm1hiz} shows the corresponding second moment 
of the occupation distribution of subhalos as a function of 
redshift.  The results are for the same model realizations 
used to generate Fig.~\ref{fig:Ncomphiz}.  The figure shows that 
unlike at $z=0$, the high redshift second moment model results are 
in good agreement with the $\lcdme$ simulation over a wide range of 
host halo masses.  As redshift increases at a fixed host mass bin, 
the masses of the host halos in that bin become increasingly 
large with respect to the typical mass of collapsing objects 
$M_{\star}$.  So as redshift increases at fixed mass, we 
are examining increasingly rare objects.  The results in 
Fig.~\ref{fig:NNm1hiz} illustrate the same qualitative trend 
that is seen in Fig.~\ref{fig:NNm1}, namely, 
that the model result for the second moment appears 
to be more like the simulation results as we examine 
increasingly rare host halos.  

%
%
%
%
\subsection{Accretion vs. Destruction} \label{sub:acc}

Although the first detailed studies of halo 
substructure showed that subhalo 
populations may scale in a simple, nearly 
self-similar way with the size of the the host 
halo \citep[e.g.][]{moore_etal99} there is 
now evidence to the contrary.  
Recently, \citet{gao_etal04a} reported the 
measurement of a deviation from the self-similar 
scaling of subhalo populations in simulated halos 
(\citeauthor{diemand_etal04} \citeyear{diemand_etal04}, 
hinted at this but did not have a large enough 
host halo sample to make a significant detection) while 
\citet{vdb_etal04} found a similar trend in subhalo 
abundance using analytic arguments.  To be specific, 
\citet{gao_etal04a} found that the number of 
subhalos with mass scaled to a fixed fraction of the 
host halo mass $\dd\Nsg/\dd(\Msat/\Mhost)$, 
increases with host mass as $\sim \Mhost^{0.1}$ in the 
regime $\Msat/\Mhost \ll 1$.  
In Figure~\ref{fig:Mfunc}, we show that the 
results of our semi-analytic model predict a 
very similar trend with host halo mass and, as 
we have a large number of realizations of host halos 
in each mass bin, the trend that we compute is 
significant.  Our model is in good agreement with 
the \citet{gao_etal04a} result, yielding a subhalo 
abundance that scales as a power-law in the 
scaled satellite mass $(\Msat/\Mhost)$, 
with a weakly host mass-dependent normalization.  
We find that 
$\dd\Nsg/\dd(\Msat/\Mhost) \propto \Mhost^{\nu} 
(\Msat/\Mhost)^{-\mu}$ in the regime 
$\Msat/\Mhost \ll 1$, with 
$\nu \approx 0.08$ and $\mu \approx 1.88$.

\begin{figure*}[!t]
\epsscale{1.8}
\plotone{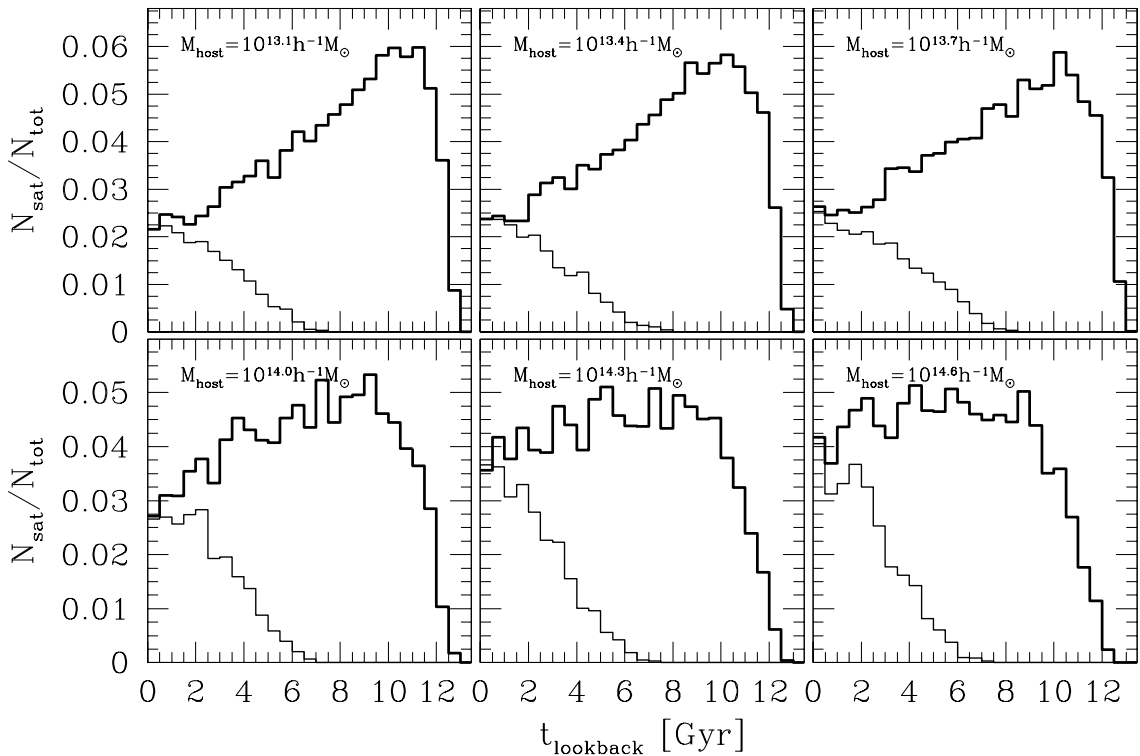}
\caption{
The distributions of subhalo accretion and 
destruction times as a function of the final $z=0$ host 
halo mass $\Mhost$.  In each panel, 
we show accretion rates for the different 
final host masses listed in the top left of the 
panel.  The {\it thick solid} upper lines indicate 
the number of all subhalos that are accreted with an 
initial maximum circular velocity 
$\vmaxsat \ge 0.2\vmaxhost$ as a function of the 
lookback time to the time of accretion 
(i.e. time {\em since} accretion) in bins of 
width $\Delta\tlookback = 0.5 \Gyr$.  
The {\it thin solid} lower lines show the number of subhalos
accreted in each lookback time bin, which remain above the 
same threshold of $\vmaxsat$ at $z=0$, after the various 
destructive effects have been accounted for.  The number of 
subhalos in each bin is shown in units of the total number of 
accreted subhalos $N_{\mathrm{tot}}$.  
}
\label{fig:dNdtVrel}
\end{figure*}

The semi-analytic model also provides a simple framework for 
understanding the nature of this trend.  The abundance of subhalos is
determined by the constant competition between subhalo accretion 
and destruction.  In Figure~\ref{fig:dNdtVrel}, we illustrate
the competition between subhalo accretion and destruction rates 
as a function of $\Mhost$ for different final ($z=0$) masses 
of the host halo.  In the standard,
$\Lambda$CDM cosmological model, the typical amplitude of mass density
fluctuations is a decreasing function of length scale (or an
associated mass scale $M = 4 \pi \rhomean R^{3} / 3$ as in
\S~\ref{sub:merger_trees}), so structure forms hierarchically as small
objects collapse early and merge to form larger objects.  
Large halos observed at redshift $z=0$ have thus typically assembled
most of their mass relatively more recently than their less massive
counterparts (e.g., see discussions in \citeauthor{lacey_cole93} and
\citeauthor{wechsler_etal02}).  Correspondingly, one important
manifestation of the more recent mass assembly of relatively massive
halos is that large host halos have typically acquired their
satellites more recently than less massive hosts.  This shift in
the relative accretion time of subhalos is evident in
Fig.~\ref{fig:dNdtVrel}.  The rate of satellite halo accretion 
(shown by the thick, upper lines) for the
$\Mhost = 10^{13.1} \hMsun$ host is strongly peaked at $\tlookback
\sim 10-11 \Gyr$ in the past and drops off rapidly at more recent
times, while for the largest, Coma-size host, with 
$\Mhost = 10^{14.6} \hMsun$ host, 
the accretion rate peaks more recently at 
$\tlookback \sim 8-9 \Gyr$ ago and remains 
nearly constant from this time until the present.  

The thin, lower lines in 
Fig.~\ref{fig:dNdtVrel} 
show the accretion rate for 
{\em subhalos that remain above the cut 
$\vmaxsat \ge 0.2\vmaxhost$ at redshift $z=0$}.  
We refer to these subhalos as 
{\em surviving} subhalos.  
In agreement with \citeauthor{zentner03}, 
\citet{gao_etal04a}, and \citet{vdb_etal04}, 
we find that only $\sim 4-12\%$ of surviving 
subhalos are accreted prior to a redshift of 
$z=1$ ($\tlookback \simeq 7.6 \Gyr$).  
At $\tlookback \sim 4 \Gyr$ 
($z \sim 0.35$), the probability of a subhalo 
surviving and remaining in the $z=0$ 
sample is roughly $50\%$ and is only very 
weakly dependent upon the mass of the host 
halo.  This timescale 
for removing subhalos from the sample due 
to mass loss and merging is 
not surprising because the typical 
dynamical times in halos at low redshift 
are of order $\sim 3-5 \Gyr$ and lends support 
to the simpler approach of \cite{vdb_etal04} 
for modeling subhalo populations in applications 
where the level of detail we present here is not 
needed.

The reason for the 
deviation from a self-similar scaling of 
subhalo populations is evident in 
Fig.~\ref{fig:dNdtVrel}.  Subhalos in larger 
hosts have typically been accreted more 
recently and have had less time to be 
tidally stripped and less time for their 
orbits to decay via dynamical friction 
and fall to the host halo center. 
The later formation times of more 
massive host halos cause the subhalo 
populations of host halos to 
deviate from self-similarity 
simply because the balance between 
typical accretion times and typical 
destruction times shifts monotonically in 
favor of accretion with increasing $\Mhost$.  
\citet{vdb_etal04} developed a similar interpretation 
of these results using an analytic model of subhalo 
populations.

%
%
%
%
%
%
%
%
%

\begin{figure*}[p]
\epsscale{1.5}
\plotone{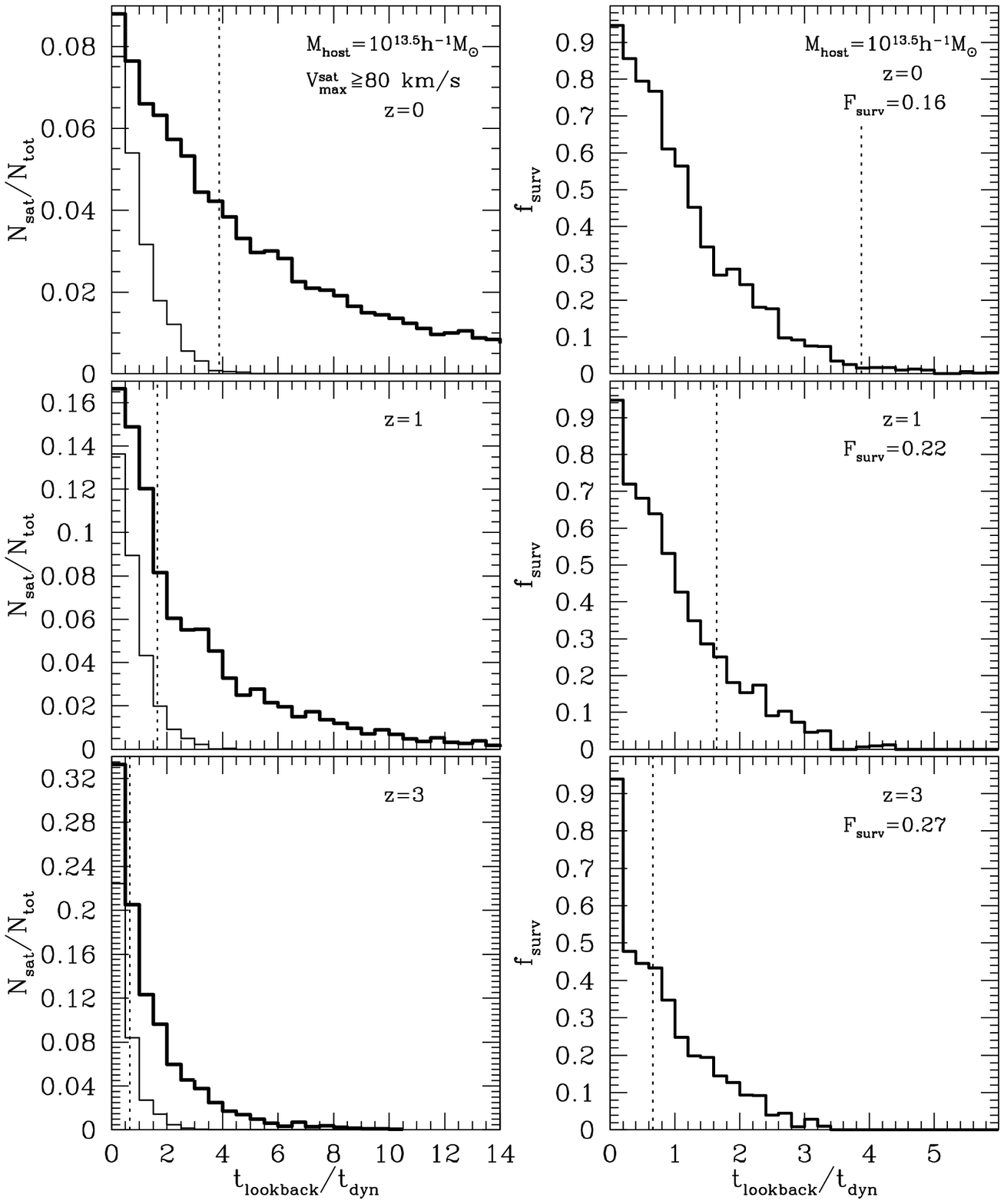}
\caption{
The competition between accretion and destruction 
at various redshifts.  
{\it Left Panels:} The {\it thick solid} lines show 
the accretion time distributions of subhalos with 
$\vmaxsat \ge 80 \kms$ at the epoch of 
accretion as a function of lookback time to 
accretion for a host of mass $\Mhost=10^{13.5} \hMsun$.  
The lookback time has been scaled by the typical 
dynamical time of the host halo at the time of 
accretion, $\tdyn$.  The {\it thin solid} 
lines show the number of accreted 
subhalos that remain above the aforementioned 
$\vmax$ threshold at the redshift of observation.  
From top to bottom, we show the model results at 
final redshifts of $z=0,1,3$.  In all panels, the 
{\it vertical dotted} line depicts the median value 
of $\tlookback/\tdyn$ for all accreted 
subhalos.    
{\it Right Panels:} The fraction of surviving 
subhalos as a function of the lookback time to 
accretion.  The lines show the fraction of 
halos that are above the threshold 
$\vmaxsat \ge 80 \kms$ 
at the time of accretion 
and remain above this threshold at the 
redshift of observation.  Again, the lookback 
time has been scaled by the dynamical time of 
the host halo at the time of accretion.  
All panels show model results for a host of mass 
$\Mhost=10^{13.5} \hMsun$ at final 
redshifts of (from top to bottom) $z=0,1,3$.  
Listed in each panel is the total number 
of subhalos that are accreted with 
$\vmaxsat \ge 80 \kms$ that remain above 
this threshold at the redshift of observation, 
$\Fsurv$.  The results shown in both panels are 
from $1000$ realizations of the  model of 
\S~\ref{sec:model}.
}
\label{fig:dNdtz}
\end{figure*}

Given the discussion of the trend in the 
normalization of subhalo mass and velocity 
functions above, it 
seems natural that the trend of subhalo 
abundance with redshift shown in 
Figure~\ref{fig:Ncomphiz} should have 
a similar explanation.  Along the 
same lines, it is also natural to 
inquire about subhalo survival probabilities 
for satellites accreted at various times.  
We address both of these 
issues in Figure~\ref{fig:dNdtz}.  
In the left hand panels of Fig.~\ref{fig:dNdtz}, 
we show the accretion rates of all subhalos 
(thick line) and of surviving halos (thin line) 
for our model realizations of host halos of 
mass $\Mhost = 10^{13.5} \hMsun$ at three 
different redshifts, $z=0,\ 1,$~and~$3$.  
In this figure, we define our subhalo samples by a 
fixed threshold of $\vmax \ge 80 \kms$.  
We refer to subhalos that remain above this threshold 
at the final redshift as {\em surviving} subhalos.  
We show accretion rates for halos at disparate redshifts 
for which the typical timescales of accretion 
and the typical dynamical timescales of halos 
are quite different from each other.  In order 
to make the results for the halos at different 
redshifts more nearly commensurable, 
in Fig.~\ref{fig:dNdtz} we assign subhalos 
an accretion time in units of the typical 
dynamical time of the host halo at the time 
of accretion $\tlookback/\tdyn$.  The scaling of 
$\tdyn$ with $\tlookback$ is responsible for the 
accretion rates dropping monotonically with time, 
as opposed to rising, as they do in 
Fig.~\ref{fig:dNdtVrel}.  In each panel, 
the vertical dotted line shows the median 
value of $\tlookback/\tdyn$ for all accreted 
subhalos.  The results are computed from $1000$ 
realizations of the model of \S~\ref{sec:model}.

Notice first in Fig.~\ref{fig:dNdtz}, 
the relative shift in the lookback time 
to accretion as a function of the redshift at 
which we observe the host halo.  
For the $z=0$ host halos, 
the median subhalo accretion time is roughly 
four halo dynamical times, while for the 
$z=3$ hosts, the median halo accretion time is 
only about one-half of a halo dynamical time.  
The reason for this shift is essentially the 
same as the reason for the shift in accretion 
times as a function of host halo mass:  large 
halos are increasingly rarer objects at high 
redshift than they are at low redshift and are 
characterized by more recent mass assembly.  

In the right-hand panels of Fig.~\ref{fig:dNdtz}, 
we explicitly show the fraction of surviving subhalos 
$f_{\mathrm{surv}}$ accreted at $\tlookback$, 
as a function of $\tlookback$ in units of the 
typical dynamical time of the host.  
In the upper right of each panel, we also show the 
total, integrated fraction of {\em all} accreted subhalos 
that survive until the final redshift, $\Fsurv$.    
In all cases, subhalos very rarely survive 
for more than $\sim 3-4$ dynamical times after they 
are accreted onto their host halos.  The 
typical timescale for destruction is roughly 
a dynamical time, as approximately $50\%$ of all 
satellite halos fall below the sample $\vmax$ 
threshold after roughly one dynamical time in each 
case.  However, notice that in the hosts observed at a 
final redshift of $z=0$, the typical time for 
a subhalo to be accreted is $\sim 3.8 \tdyn$, 
while for the host observed at $z=3$, the 
typical subhalo accretion time is only 
$\sim 0.6 \tdyn$.  This shift in the 
accretion times relative to the timescale that is 
relevant for removing a halo from the 
threshold sample is dramatic and is responsible for 
the increase in the overall subhalo survival 
fraction as $\Fsurv = 0.16,\ 0.22,$~and~$0.26$ for the 
redshifts $z=0,\ 1,$~and~$3$ halos respectively, and 
it is the reason for the systematic increase in the 
abundance of subhalos in fixed mass hosts with 
increasing redshift illustrated in Fig.~\ref{fig:Ncomphiz}.

%
%
%
%
%
%
%
%

%
%
%
%
%
\subsection{Host Properties and Subhalo Populations}
\label{sub:hostsub}
%
%

The discussion of the previous section has a natural extension.  
As can be seen in Fig.~\ref{fig:Vfunc}, there is a 
sizable scatter in the number of subhalos of a 
particular size in a host of {\em fixed} mass.  In 
the preceding paragraphs, we demonstrated that 
more massive host halos typically contain more 
substructure simply because they assembled their mass 
more recently and their subunits have had less time to 
evolve dynamically.  
A natural implication of this is that the scatter in 
the amount of halo substructure at fixed host halo mass 
may be largely determined by the variety of mass 
accretion histories of halos at fixed mass:  halos that 
acquired their mass more recently should have more 
substructure.  

%
%
%
%
\begin{figure*}[t]
\epsscale{1.75}
\plotone{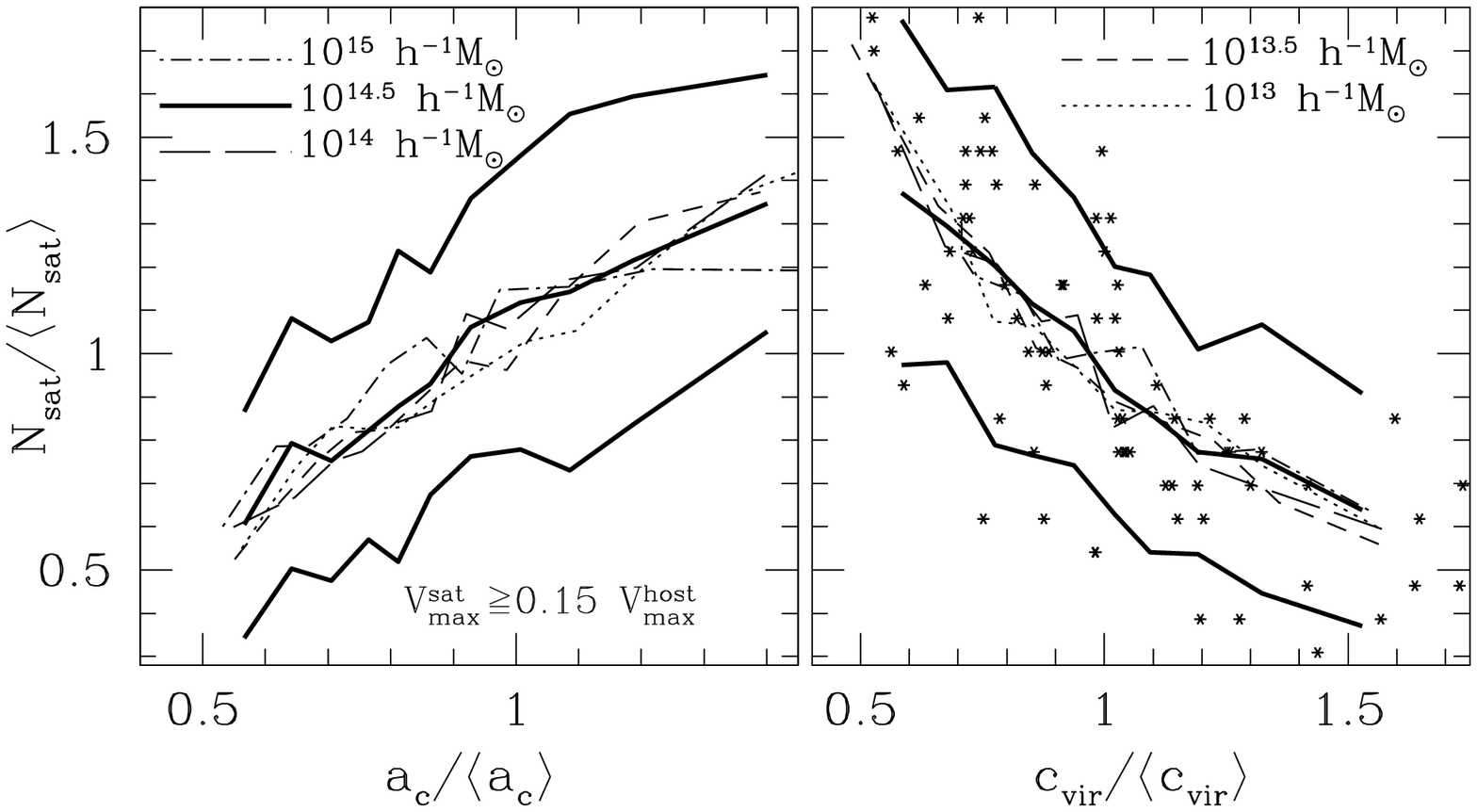}
\caption{
{\it Left Panel:}  Correlation between the 
collapse epoch as defined in Eq.~(\ref{eq:Mofa}) and the 
number of surviving subhalos.  We plot the mean relation 
between the halo collapse epoch $\ac$, scaled by the mean collapse 
epoch $\langle \ac \rangle$, and the number of satellites $\Nsg$, 
scaled by the mean number of satellites $\langle \Nsg \rangle$, 
predicted by $1000$ realizations of our substructure model.  The 
central {\it thick solid} line corresponds to the mean relation 
for $\Mhost=10^{14.5} \hMsun$ hosts.  The other line types represent 
the mean relation at four different host masses:  
$10^{15} \hMsun$ ({\it dot-dashed}), $10^{14} \hMsun$ ({\it long-dashed}), 
$10^{13.5} \hMsun$ ({\it short-dashed}), and 
$10^{13} \hMsun$ ({\it dotted}).  
The upper and lower {\it thick, solid} lines delimit the 
realization-to-realization scatter in the relation based 
on the $1000$ model realizations of a $\Mhost = 10^{14.5} \hMsun$ host.  
We count only subhalos with $\vmaxsat \ge 0.15 \vmaxhost$ at $z=0$.  
{\it Right Panel:}  The relationship between host halo concentration 
and the number of satellite halos.  The lines represent the mean 
relation between halo $\cvir$ scaled by the mean concentration 
$\langle \cvir \rangle$, and the number of satellite halos scaled 
by the mean number of satellite halos $\Nsat$, predicted by $1000$ 
model realizations.  The different line types correspond to different 
host masses and are the same as in the left panel.  
Again, only subhalos with $\vmaxsat \ge 0.15 \vmaxhost$ are counted.  
The stars represent the measured concentrations and satellite 
numbers for all host halos in the $\lcdme$ simulation with 
$\vmaxhost \ge 500 \kms$.  A power-law fit to the simulation 
points yields $\Nsg \propto \cvir^{-a}$ with 
$a \approx 0.98 \pm 0.15$.  
The relationship from the subhalo model yields a slightly 
smaller best-fit power-law index of 
$a \approx 0.86 \pm 0.07$. 
}
\label{fig:ACNsat}
\end{figure*}
%
%
%
%

A convenient quantity that can be used to describe the 
variety of mass assembly histories of halos is the 
collapse epoch $\ac$, introduced by \citeauthor{wechsler_etal02} 
and defined in Eq.~(\ref{eq:Mofa}) above.  As we described in 
\S~\ref{sub:concentrations}, for each realization of our 
model, we determine the collapse epoch of the host halo by 
fitting Eq.~(\ref{eq:Mofa}) to its mass assembly history.  

In the left panel of Figure~\ref{fig:ACNsat}, we show the correlation
between $\ac$ and satellite number predicted by our model for five
different values of host mass.  For each $\Mhost$ value, we plot the
number of satellites with $\vmaxsat \ge 0.15 \vmaxhost$, in units of
the mean number of satellites for hosts of this mass, as a function of
$\ac$ in units of the mean collapse epoch $\langle \ac \rangle$, for
hosts of this mass.  The thick, solid, central line corresponds to the
mean relation for $\Mhost = 10^{14.5} \hMsun$ hosts and the upper and
lower solid lines show the scatter in this relation measured from
$1000$ model realizations.  The remaining lines show the mean relation
between collapse epoch and satellite number for the four different
host halo masses indicated in the Figure caption.  The strong correlation
between collapse epoch and satellite number is apparent in
Fig.~\ref{fig:ACNsat}, with host halos that have an average collapse
epoch hosting an average number of satellite halos.  We see no
evidence that this relation changes significantly as a function of
host mass.  Additionally, at fixed $\Mhost$ and $\ac$, there remains a
significant amount of scatter in the number of satellite halos,
indicating that there are other important physical ingredients that
affect the amount of halo substructure.

In the context of our model, two possible sources of this scatter are:
(1) the distribution of orbital parameters and subhalo concentrations
for a fixed host halo accretion history; (2) the variety of accretion
histories that result in the same best-fit $\ac$.  By constructing
$100$ model realizations using a {\em single} host mass accretion
history in several mass bins, we estimate that approximately half of
the scatter can be attributed to (1) above.  

As we mentioned in \S~\ref{sub:concentrations}, the concentrations of
halos correlate strongly with their mass accretion histories.  In
fact, we used this correlation to set the concentrations of our host
halos according to Eq.~(\ref{eq:cofac}), as prescribed by
\citeauthor{wechsler_etal02}.  The implication is that the fundamental
correlation between mass accretion history and satellite abundance
implies a correlation between host $\cvir$ and $\Nsg$, with $\Nsg$ a
decreasing function of host halo concentration.  In the context of our
model, this correlation must be present given the correlation between
$\ac$ and $\Nsg$ shown in the left panel of Fig.~\ref{fig:ACNsat} and
the fact that we assign concentrations via Eq.~(\ref{eq:cofac}).  We
illustrate the model results for the $\cvir$-$\Nsg$ relation in the
right hand panel of Fig.~\ref{fig:ACNsat}.  Again, the different lines
correspond to the model predictions at various host halo masses.  The
strong correlation between halo concentration and satellite number is
clear.  For this subhalo selection criterion a power-law fit to the
semi-analytic model result yields $\Nsg \propto \cvir^{-a}$ with $a
\approx 0.86 \pm 0.07$.

It is interesting to test the extent to which the simulated halos
exhibit this correlation between concentration and subhalo number.
The {\em stars} in Fig.~\ref{fig:ACNsat} show the relationship between
$\Nsg$ and $\cvir$ for host halos in the $\lcdme$ simulation.  The
values for the simulation halos were computed by first selecting host
halos with $\vmaxhost \ge 500 \kms$ in order to ensure the
completeness of the subhalo count down to a threshold of $\vmaxsat \ge
0.15\vmaxhost$ (considering satellites above the scaled threshold
roughly eliminates the scaling of satellite halo number with host halo
size).  We then fit the host density profile to an NFW profile
[Eq.~(\ref{eq:nfw_profile})] in order to determine the best-fit
$\cvir$.  Finally, we used the subhalo counts and best-fit
concentrations to compute $\Nsat$ and $\langle \cvir \rangle$.  The
simulated host halos also show a clear correlation between
concentration and satellite number.  The small number of host halos in
this sample makes it difficult to draw detailed conclusions, yet the
$\cvir$-$\Nsg$ correlation measured in the simulation appears to be in
good agreement with the model predictions.  This correlation also
appears in agreement with the related correlation between halo $\vmax$
and satellite number observed in the simulations of
\citet{gao_etal04a}.  Fitting the simulated halo results to a power
law $\Nsg \propto \cvir^{-a}$ yields a slightly steeper relation than
the model fit, with a best-fit power-law index of $a \approx 0.98 \pm
0.15$.  Given the errors, the power-law indices from the model and
simulation data fits are consistent with each other.  
Interestingly, we find no statistically significant 
trend in the radial distribution of satellites with 
host halo concentration or collapse epoch.  

Figure~\ref{fig:ACNsat} shows the correlation between 
$\ac$ and the number of satellite halos selected according 
to a specific selection threshold, namely satellites with 
$\vmaxsat/\vmaxhost \ge 0.15$.  However, we expect the 
strength of the correlation to be dependent upon the satellite 
sample threshold for several reasons.  The drag 
of dynamical friction is proportional to $\Msat^2$, so dynamical 
friction is most efficient at driving rather massive satellites 
toward the centers of their hosts.  Furthermore, the median 
lookback time to a merger with a satellite of a 
specific mass is a decreasing function of $\Msat$.  
Mergers with large subhalos occur relatively more recently, 
on average, than mergers with small subhalos. 
In addition, the accretion of large subhalos dominates 
the rate of host mass growth, so the accretion times of 
relatively massive subhalos more strongly influence any 
particular definition of formation epoch than the accretion 
times of small subhalos.  These factors 
suggest that the strength of the $\ac$-$\Nsg$ (or $\cvir$-$\Nsg$) 
correlation should depend upon the size of the subhalos 
included in the sample and that the number of 
large subhalos should be more strongly dependent upon 
the mass accretion history of the host than the number 
of small subhalos.  It is possible to use our substructure 
model to explore this threshold dependence, because our model 
provides a large number or realizations of host halos 
of various masses so that we are not hampered by the limited 
dynamic range of the cosmological simulation.

\begin{figure}[t]
\epsscale{1.0}
\plotone{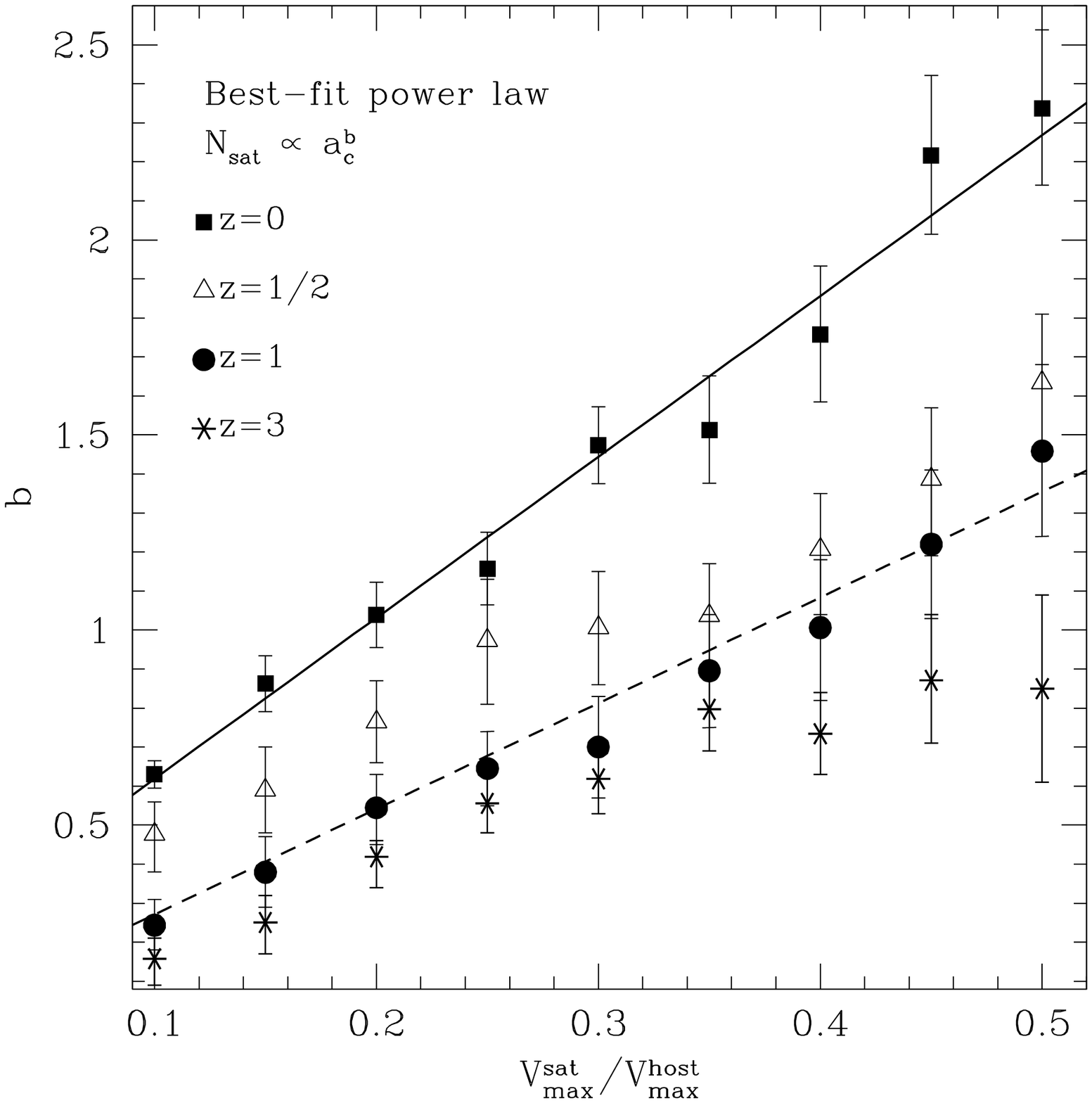}
\caption{
The $\Nsg$-$\ac$ relationship as a function of subhalo 
$\vmax$ threshold.  We show the best-fitting power law indices 
from a fit of the number of satellites above a particular $\vmax$ threshold 
$\Nsg(>\vmaxsat/\vmaxhost)$, to a power law function of the 
host halo collapse epoch $\Nsg(>\vmaxsat/\vmaxhost) \propto \ac^b$, 
as a function of the satellite threshold $\vmaxsat/\vmaxhost$ 
at redshifts $z=0$ ({\em squares}), $z=1/2$ ({\em triangles}), 
$z=1$ ({\em circles}), and $z=3$ ({\em stars}).  The {\it solid} 
line represents a fit of the power law index $b$, as a function 
of threshold to a line, $b \simeq 0.25 + 4(\vmaxsat/\vmaxhost)$ at 
$z=0$.  The {\it dashed} line represents a linear fit at $z=1$, 
$b = 2.7(\vmaxsat/\vmaxhost)$.
}
\label{fig:acnpl}
\end{figure}

We demonstrate the $\vmax$ threshold dependence of the 
relation between $\Nsg$ and $\ac$ in Figure~\ref{fig:acnpl}.  
The abscissa represents minimum 
values of scaled maximum circular velocity $\vmaxsat/\vmaxhost$, 
used to define various subhalo samples.  At each threshold, 
we fit the number of satellite halos above the threshold 
to a power law 

\beq
\label{eq:Nsgfit}
\Nsg(>\vmaxsat/\vmaxhost) \propto \ac^b.
\eeq
At all thresholds, we find a power law to be an acceptable 
fit and plot the value of the best-fit power law indices on 
the vertical axis.  The black squares show the relationship 
between $b$ and $(\vmaxsat/\vmaxhost)$ at $z=0$.  
That the strength of the correlation 
between satellite number and $\ac$ is a strong 
function of the $\vmax$ threshold is 
evident in the Figure~\ref{fig:acnpl}.  The solid line  
shows the linear relation
\beq
\label{eq:bfit}
b = 0.25 + 4(\vmaxsat/\vmaxhost),
\eeq
which we find to be a good fit to the relationship between 
the power law index $b$, and the sample threshold 
$\vmaxsat/\vmaxhost$, in the regime 
$0.08 \le \vmaxsat/\vmaxhost \le 0.5$ at $z=0$.  
As can be seen 
in Fig.~\ref{fig:acnpl}, the strength of the $\Nsg$-$\ac$ 
correlation declines with increasing redshift as dynamical 
processes become less influential in determining subhalo 
populations.  At $z=1$, the best-fit relation for the power 
law index is $b \simeq 2.7(\vmaxsat/\vmaxhost)$.  The 
marked evolution between redshift $z=0$ and $z\approx 1$ is 
due to the cosmological constant (recall $\Omegal = 0.7$) 
which causes a dramatic reduction in the rate of structure 
growth, and therefore satellite accretion rates, at low 
redshift.

%
%
%
%
%
%
%
%

\begin{figure}[t]
\epsscale{1.0}
\plotone{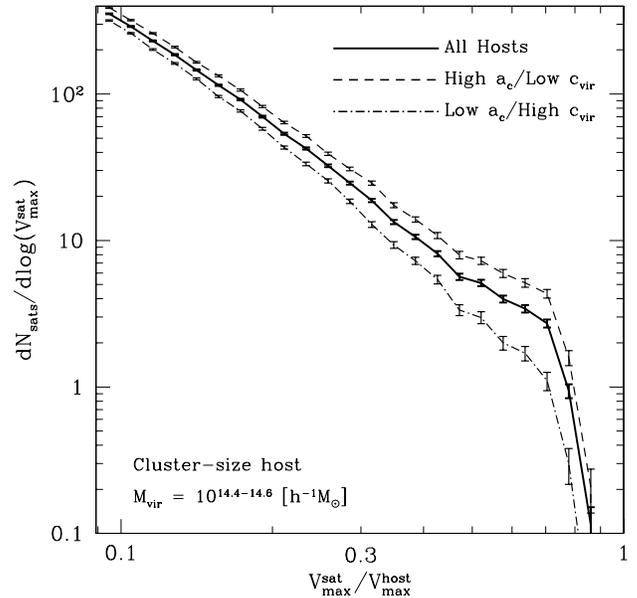}
\caption{
The dependence of the differential velocity function on 
host halo formation time and concentration.  We show the 
differential velocity functions of subhalos in cluster-size 
hosts computed by stacking the $1000$ model realizations 
in three host mass bins, 
$\Mvir = 10^{14.4}$,~$10^{14.5}$,~and~$10^{14.6} \hMsun$.  
The {\em thick, solid} central line represents the average 
differential velocity function for all $3000$ of the host halo 
realizations.  The {\em dashed} line depicts the mean differential 
velocity function for the half of the host halo sample with the 
highest $\ac$ (lowest $\cvir$) and the {\em lower, dot-dashed} line 
represents the differential velocity function for the 
half of the host halo sample with the lowest $\ac$ 
(highest $\cvir$).  In all cases, the error bars represent the 
estimated error on the mean value.  
}
\label{fig:dndlgv}
\end{figure}

Another representation of the 
detailed dependence of the correlation between 
satellite abundance and host halo formation epoch 
(or host halo concentration) on the relative subhalo 
size is displayed in Figure~\ref{fig:dndlgv} 
for cluster-size host halos.  
The Figure shows the differential velocity 
functions (DVF) of subhalos as a function 
$\vmaxsat/\vmaxhost$ for all host halos and for two 
subsamples subdivided by host halo formation epoch.  
The curves in Fig.~\ref{fig:dndlgv} were constructed by 
stacking our $1000$ model realizations of hosts at three 
masses, $\Mvir = 10^{14.4}$,~$10^{14.5}$,~and~$10^{14.6} \hMsun$, 
for a total of $3000$ realizations, 
in order to overcome noise in the measurement.  
The solid line represents the mean DVF for all $3000$ 
host halos.  The upper, dashed line represents the DVF 
for the half of the host halo sample with the highest 
$\ac$ (lowest $\cvir$), while the lower, dot-dashed line 
represents the DVF for the half of the sample with the 
lowest $\ac$.  Again, the dependence of subhalo populations 
on host halo accretion history is evident.  Late-forming, 
low concentration host halos have more substructure of 
all sizes and larger subhalos are more strongly correlated 
with accretion history and host halo $\cvir$ for the 
aforementioned reasons.  We speculate on possible implications 
of these relationships in \S~\ref{sec:conclusions}.

%
%
%
%

%
%

%
%

\subsection{Application to Galaxy Clustering}\label{sub:app}

The model that we have presented has the virtues 
of being simple, quick and easy to compute, 
easy to parse and understand, and easy 
to modify or add specific ingredients to.  
These features make the model 
useful for studying a wide range of phenomena.  
In this section, we briefly discuss the particular 
example of applying this model to make predictions for 
the two-point correlation function of galaxies. 
This subject will be considered in much greater detail
in the subsequent papers in subsequent papers of this 
series.

\begin{figure}[t]
\epsscale{1.0}
\plotone{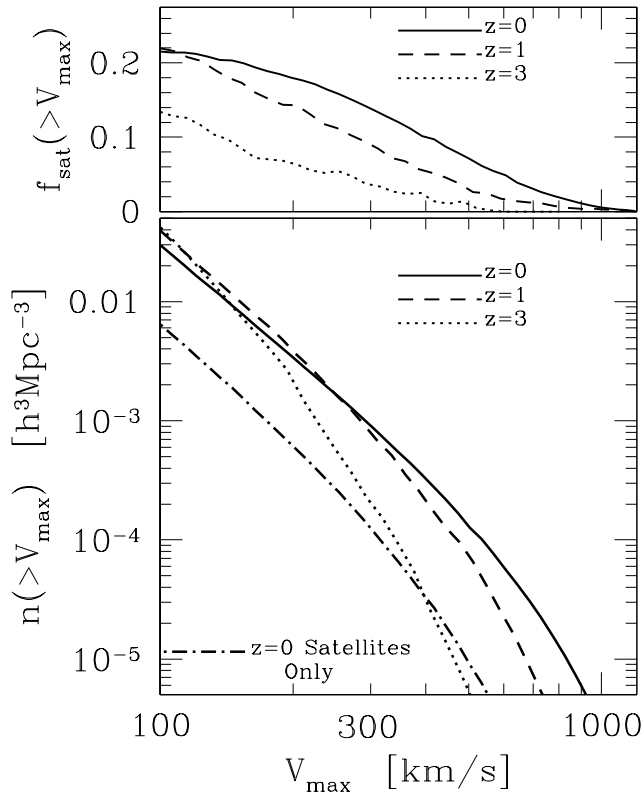}
\caption{Halo and subhalo populations at three redshifts.  
{\it Bottom Panel}:  The total number of halos 
(host halos {\em and} subhalos) with a maximum circular velocity 
greater than $\vmax$ as a function of $\vmax$.  The {\it solid} line 
shows the halo population at $z=0$, the {\it dashed} line shows the 
halo population at $z=1$ and the dotted line shows the halo population 
at $z=3$.  The {\it dot-dashed} line shows the contribution from 
subhalos only at $z=0$.  {\it Top Panel}:  The fractional contribution of 
subhalos to the total number of halos above each $\vmax$ threshold 
at redshifts $z=0,1,3$.  The linetypes correspond to the same redshifts 
as in the {\it bottom} panel. 
}
\label{fig:Fsats}
\end{figure}

We begin by using our model to compute the abundance of subhalos in a
cosmological volume.  By convolving the mean of the subhalo 
occupation distribution shown in Fig.~\ref{fig:Ncomp} and 
Fig.~\ref{fig:Ncomphiz}, with the known mass function of host halos
\citep{jenkins_etal01}, we can compute the number density of all
halos, including both host halos and their subhalos, as a function of
$\vmax$ threshold.  The result is shown in the bottom panel of
Figure~\ref{fig:Fsats} at three redshifts: $z=0$,~$1$, and $3$.  The
decrease in number density as a function of redshift is a reflection
of the fact that fewer massive halos have collapsed at earlier epochs.
The dot-dashed line in the lower panel of Fig.~\ref{fig:Fsats}
corresponds to the mean number density of satellite halos only at
$z=0$.  In the upper panel of Fig.~\ref{fig:Fsats}, we show the
fraction of halos that are subhalos as a function of $\vmax$
threshold, $\fsat(>\vmax)$.  At low redshift, $\fsat$ is roughly $\sim
16-22\%$ at thresholds that correspond to galaxy-size halos, which is
comparable to the fraction of galaxies observed to be in groups and
clusters.  The decrease in $\fsat$ with increasing $\vmax$ threshold
reflects the fact that the mass function of host halos 
$\dd n/\dd \Mhost$, 
is an increasingly steeply declining function of host halo
mass.  Therefore, as the threshold is increased, the number density of
host halos becomes increasingly dominated by host halos near the
$\vmax$ threshold.  Halos are very unlikely to host satellites with a
comparable $\vmax$ (see Fig.~\ref{fig:Vfunc}), so $\fsat$ decreases
monotonically with increasing $\vmax$.  A qualitatively similar
argument applies to the dependence of $\fsat$ on redshift at fixed
$\vmax$. The mass functions and the fraction of satellites are 
in good agreement with results of simulations 
presented by \citet{kravtsov04a}.

%
%
%
%
%
%
%
%

In \S~\ref{sec:intro}, we suggested that our model of substructure 
could be coupled with a low-resolution, large-volume simulation or 
with an analytic halo model \citep[e.g.][]{seljak00} 
in order to study the clustering properties of galaxies.  We 
pursue these applications further in the forthcoming papers of this 
series.  However, the fact that the subhalo populations 
computed from our model are in good agreement with the subhalo 
populations of halos observed in cosmological $N$-body simulations 
does not imply that statistics that describe the spatial clustering 
of halos, like the two-point correlation function $\xi(r)$, 
will be in good agreement.  

Consider first, populating the host halos of a 
low-resolution simulation with subhalo populations 
derived from the model of \S~\ref{sec:model}.  
In the most common implementation of the EPS 
formalism for hierarchical clustering 
(\citeauthor{bond91} \citeyear{bond91}; \citeauthor{lacey_cole93}; 
\citeauthor{somerville_kolatt99} \citeyear{somerville_kolatt99}), 
which we have used to generate mass accretion histories for our 
host halos, the accretion history of a halo at fixed mass is 
independent of its large-scale environment.  There is some evidence 
suggesting that this practical approximation may be justified 
(\citeauthor{lemson_kauffman99} \citeyear{lemson_kauffman99}; 
but see also 
\citeauthor{sheth_tormen04} \citeyear{sheth_tormen04}).
Yet, significant correlations between environment and host halo 
properties, such as $\ac$ or, more generally, 
the entire halo mass accretion history, would also imply 
correlations between environment and subhalo populations.  
These correlations would then result in systematic differences 
between correlation functions measured directly from a simulated 
halo sample and those measured from a sample constructed 
by populating simulated hosts with subhalos from our model 
and it is important to understand this potential systematic 
effect before applying this method.  As a byproduct, 
we also test the extent to which environmental effects may 
influence the correlation function and one of the fundamental 
assumptions of the halo model, the statistical independence 
of halo properties from halo environment.

We attempted to estimate the potential importance of these effects
using the following procedure.  We identified host halos in the
$\lcdme$ simulation as described in \S~\ref{sec:simulations}, and
computed their virial masses and the positions of their most bound
particles.  We then used these masses and positions as inputs to our
substructure model.  We computed a semi-analytic mass accretion
history and a corresponding subhalo population for each host in the
simulation, thereby constructing a hybrid catalog of simulated host
halos and semi-analytic model subhalos.  We reiterate that we did {\em
  not} use the concentrations or $\vmax$ values of the simulated hosts
in order to construct the mass accretion histories and subhalo
populations.  Instead, we assigned each host halo new values of
$\cvir$, $\vmax$, and $\ac$ based on the semi-analytic mass accretion
histories that we generated and the method outlined in
\S~\ref{sub:concentrations}.  As such, these host properties and the
properties of their subhalos are re-assigned in a way that is
independent of any effects due to the large-scale environment.
However, we did use the positions of the hosts to construct our hybrid
catalogs, so the hybrid catalogs exhibit the same large-scale
structures as the simulation catalogs and there should thus be no
differences between the two due to cosmic variance.  Within each
simulated host halo, we place our model subhalos at a distance
$R/\Rvir$ from the halo's most bound particle, where $R/\Rvir$ is
given by the dynamical model.  The subhalos are thus forced to have
spherical symmetry within their hosts and do not trace the radial
profiles of the N-body halos they occupy.  We repeated this procedure
ten times for each host halo in the $\lcdme$ simulation, thereby
generating ten realizations of independent hybrid halo catalogs in
order to account for variation from realization-to-realization of the
semi-analytic model.  Finally, we computed two-point correlation
functions for the halos+subhalos in the hybrid catalogs and the
$\lcdme$ simulation.

\begin{figure}[t]
\epsscale{1.05}
\plotone{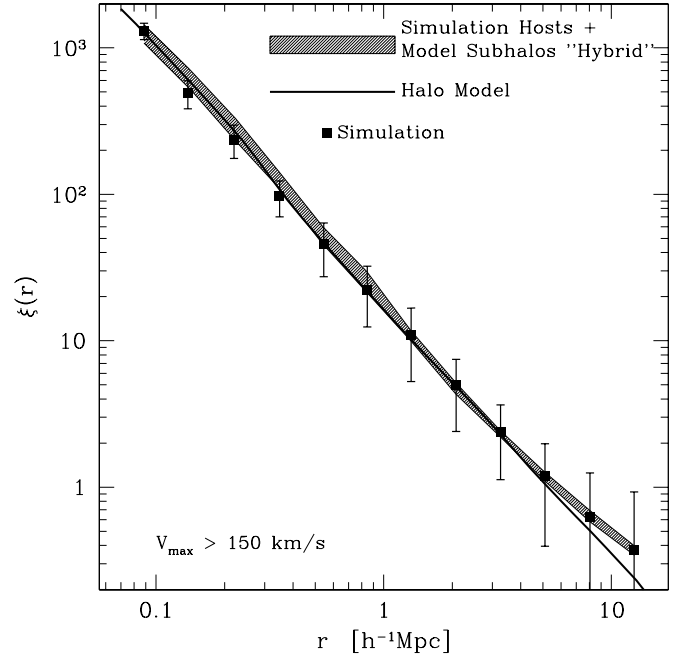}
\caption{
Halo two-point correlation functions for all halos 
(host halos and subhalos) with $\vmax \ge 150 \kms $.  
The {\it squares} represent the correlation function measured 
from the halos in the $\lcdme$ simulation.  
The error bars on the points represent the maximum of the 
jackknife error bars computed from the octants of the simulation 
box and the Poisson error based on the number of pairs.  
The {\it shaded} region represents the envelope spanned by 
ten realizations of our substructure model for each 
host halo in the simulation 
(see text \S~\ref{sub:app} for details).  
The {\it solid} line represents the 
correlation function computed by using the halo occupation 
distributions in Fig.~\ref{fig:Ncomp} and Fig.~\ref{fig:NNm1}, 
in a purely analytic computation based on the halo model 
of clustering.
}
\label{fig:CF}
\end{figure}

We summarize the results of this experiment in Figure~\ref{fig:CF},
where we show the correlation functions for samples selected above a
$\vmax$ threshold of $\vmax \ge 150 \kms$.  As a reference, 
this threshold corresponds to a mean number 
density of $n \approx 8 \times 10^{-2} \hvol$ 
(see Fig.~\ref{fig:Fsats}), which is close to the observed number 
density of galaxies with $r$-band absolute 
magnitudes of $M_r \lsim -19.5$ \citep{blanton_etal03}.  
The squares in Figure~\ref{fig:CF} represent
the correlation function of the halos in the $\lcdme$ simulation. The
error bars are the maximum of the jackknife error computed
by excluding, in turn, each of the eight octants of the simulation
volume and the Poisson error based on the number of pairs at each
separation.  A power-law fit to these points yields $\xi(r) \simeq
(r/5 \hMpc)^{-1.76}$.

The shaded band shown in Fig.~\ref{fig:CF} shows the 
envelope of the ten correlations functions computed 
from our ten hybrid halo catalogs with host halo masses 
and positions taken from the $\lcdme$ simulation and 
all other halo properties, particularly the subhalo 
populations, determined using the semi-analytic model.  
The correlation functions of the halos of our hybrid 
model calculations are generally in remarkably good 
agreement with the correlation function of the 
simulated halos.  On intermediate separations, 
$r \sim 0.5-1 \hMpc$, $\xi(r)$ of the hybrid model 
halos is slightly higher ($\sim 15\%$) than the 
correlation function of the $\lcdme$ halos. 
These separations correspond to a scale where 
most halo pairs are from halos that reside within 
the same host halo (including the host itself if 
it is part of the sample).  
In fact, the number of halo pairs on these  
scales is dominated by pairs within a common host 
in the mass range 
$10^{13} \lsim \Mhost/\hMsun \lsim 10^{14}$ 
\citep[for example, see][]{berlind02}.  
The number of pairs within a single host halo is 
$N(N-1)/2$, where $N$ is the number of halos within 
the host and $N = 1 + \Nsg$ if the host is included 
in the sample while $N = \Nsg$ otherwise.  Schematically 
then, $\xi(r)$ on separations $r \sim 0.5-1 \hMpc$ is 
roughly set by $\langle N(N-1)\rangle$ for hosts with 
$10^{13} \lsim \Mhost/\hMsun \lsim 10^{14}$.  The 
slight enhancement of $\xi(r)$ on these scales 
for the hybrid model is consistent with the differences 
in the second moment of the subhalo occupancy between 
the model results and the $\lcdme$ subhalos shown in 
Fig.~\ref{fig:NNm1}.

The fact that the two-point function of the hybrid model 
halos is in good agreement with the two-point function 
of the simulated $\lcdme$ halos is encouraging.  First, 
the properties of our hybrid model halos are independent 
of environment.  Therefore, this calculation provides 
an explicit demonstration that any correlations of 
host halo properties, such as $\cvir$, $\ac$, 
$\Nsg(>\vmaxsat)$, or the spatial distribution of subhalos 
within their hosts, with environment are 
sufficiently weak as to be nearly unmeasurable in $\xi(r)$.  
This implies that populating comparably low-resolution, 
large-volume simulations with subhalos from our model is a 
viable method for studying the two-point statistics of 
galaxies, though one may likely use a more sophisticated 
mapping of galaxies onto halos and subhalos than the simple 
$\vmax$ threshold that we used to test the importance of 
environmental effects.  We explore such models in a 
forthcoming paper.  

Additionally, the halo model of galaxy clustering 
\citep{neyman_scott52, scherrer_bertschinger91, 
seljak00, scoccimarro_etal01}
has been widely used recently for a variety of applications, 
one of which is to model galaxy clustering properties and 
to infer the galaxy occupation of dark matter halos from 
observational data 
\citep[e.g.][]{zehavi_etal03,zehavi_etal04,zheng04}.  
One of the fundamental assumptions of the halo model is that 
the galaxy occupation of dark matter halos is statistically 
independent of host halo environment and depends {\em only} 
upon $\Mhost$.  Our results indicate that this is likely an 
acceptable assumption that does not lead to considerable 
systematic errors in $\xi(r)$.  We close this section by 
illustrating the utility of our model with an explicit, 
entirely analytic calculation of the two point function 
of halos using the subhalo occupation distributions depicted in 
Fig.~\ref{fig:Ncomp} and Fig.~\ref{fig:Ncomphiz}, coupled 
with the halo model.  We compute the analytic correlation 
function as described in detail in \citet{zehavi_etal04}, 
using the host halo mass function of \citet{jenkins_etal01}, 
the dark matter correlation function of \citet{smith_etal03}, 
and the scale-dependent halo bias model of 
Tinker et al. (2004, in preparation).  We account for the 
radial profile of subhalos in our model by fitting in each 
host mass bin, the mean radial distribution of subhalos 
to a profile with a constant density core: 
$n_{sat}(r) \propto (1+r/r_{\mathrm{c}})^{-3}$.  
The resulting correlation function is shown as the solid 
line in Fig.~\ref{fig:CF}.  The agreement between all three 
methods of computing $\xi(r)$ is apparent at all separations 
in the Fig.~\ref{fig:CF} and is quite impressive.  
The fact that the halo model calculation is 
lower on large scales, $r \gsim 8 \hMpc$, is due 
to cosmic variance resulting from the finite size of the 
simulation box ($L = 80 \hMpc$).

%
%
%
%
%
%
%
%
%

%
%

\section{Discussion and Conclusions} \label{sec:conclusions}

In this paper, we presented a semi-analytic 
model for subhalo populations within host 
dark matter halos.  The details of the model are 
discussed in \S~\ref{sec:model}, with some complementary 
information in \citeauthor{zentner03}.  
In the context of this model, the evolution of a 
subhalo population is treated as a multi-stage 
process and the transitions between each stage 
are treated as abrupt.  The main ingredients of 
this model are as follows:  

\begin{itemize}

\item[1.] Semi-analytic halo merger histories, computed 
using the EPS formalism;

\item[2.] A method for assigning halo density profiles that 
incorporates the known correlations between mass accretion 
history and halo structure;

\item[3.] A distribution of initial 
orbital energies and angular momenta for merging subhalos;

\item[4.] A prescription for modeling the orbital evolution 
of subhalos within hosts, including the effects of orbital 
decay via dynamical friction and tidal mass loss.

\end{itemize}

In \S~\ref{sec:results}, 
we make a detailed comparison between the results of 
of our model and the results of cosmological $N$-body 
simulations in the regime where they can be compared 
robustly.  We calibrated our one-parameter model to 
match the normalization of the mean cumulative velocity 
function of subhalos within simulated, MW-size 
($\Mhost \approx 10^{12} \hMsun$) host halos at $z=0$, 
in a standard $\Lambda$CDM cosmology with a power spectrum 
normalization of $\sigma_8 = 0.9$.  Subsequently, 
we found, in the range that the simulation predictions 
are robust, our model to be in good agreement with 
the results of direct simulation.  We summarize the 
success of our model when compared to simulation 
results with the following points.  

\begin{itemize}

\item[1.] The velocity functions of cluster-size halos in a 
$\Lambda$CDM cosmology with $\Omegam=1-\Omegal=0.3$, 
$h=0.7$, and $\sigma_8=0.9$ are in good agreement.

\item[2.] We find that the mass functions of MW-size and 
cluster size halos in the same, 
$\Lambda$CDM cosmology are also in agreement.

\item[3.] The model predicts a radial distribution 
of satellite halos within their hosts that is in 
good agreement with simulation results and exhibits a 
sizable anti-bias of subhalos with respect to the 
dark matter.  This is in contrast to the recent 
results of \citet{taylor_babul04c}.

\item[4.] We find a good agreement between the 
first and second moments of the halo occupation 
distribution predicted by the model
and measured directly in cosmological simulations. 
The agreement holds for a wide range of host masses 
from $\Mhost \approx 5 \times 10^{11} \hMsun$ to 
$\Mhost \approx 3 \times 10^{14} \hMsun$ at redshifts 
$z=0$,~$1$~and~$3$ in a $\Lambda$CDM cosmology with a 
normalization of $\sigma_8=0.75$, 
{\it different} from the normalization
of the simulations used to provide inputs and to 
calibrate the model.

\item[5.] We find a correlation between host halo 
collapse epoch $\ac$, and satellite number $\Nsg$, 
at fixed host halo mass in our model.  
This implies a corresponding correlation 
between host $\cvir$ and $\Nsg$ at fixed $\Mvir$.  
We find that $\Nsg \propto \cvir^{-a}$, with $a \simeq 0.86 \pm 0.07$ 
for subhalos selected such that $\vmaxsat/\vmaxhost > 0.15$.  
In the regime where this is testable with the 
simulations, we find a similar correlation.  
A fit to the simulation results yield a power 
law index $a \simeq 0.98 \pm 0.15$.  Our model 
results suggest that this power law index varies 
rapidly with the subhalo selection criterion.

\item[6.] Our model results in deviations from 
a self-similar scaling of subhalo abundance with host 
halo mass.  In particular, we find that the number of 
satellite halos at a fixed value of $\Msat/\Mhost$ 
scales in proportion to $\sim \Mhost^{0.08}$, in 
approximate agreement with the recent numerical study 
of \citet{gao_etal04a} and the analytic work of 
\citet{vdb_etal04}.

\end{itemize}

The success of our model in matching the results 
of simulations in a variety of ways and at a variety of 
redshifts is non-trivial.  It is also encouraging.  
This success suggests that such a model can be 
used for a wide range of applications.  
However, note that in contrast to our results, 
\cite{taylor_etal03,taylor_babul04c}, 
find that their analytic subhalo distributions are 
significantly more centrally concentrated than 
simulated subhalo populations.  The two models 
have several different ingredients and the root 
of this discord is not yet clear.  Our results 
indicate that a centrally concentrated subhalo 
distribution is not a generic prediction of 
semi-analytic models.  

We also pointed out one possible shortcoming of the model.  Although
the semi-analytic procedure reproduces the mean quantities observed in
simulations quite well, it seems that at least over some range of host
properties, the model over-predicts the second moment of the subhalo
occupation $\langle \Nsg(\Nsg-1)\rangle$ when $\Nsat$ is less than a
few (see Fig.~\ref{fig:NNm1} and Fig.~\ref{fig:alpha}).  The model of 
\citet{vdb_etal04} exhibits this same feature notwithstanding the fact 
that their treatment of mass loss is quite different from ours.  
We speculate that the origin of this discrepancy is a fundamental 
shortcoming in the standard EPS formalism for generating 
halo merger trees, although further work is necessary to confirm 
this hypothesis.  \citeauthor{wechsler_etal02} found that EPS merger 
histories have higher scatter in formation times than simulated 
merger histories \citep*[see also][]{lin_etal03}; 
given the relation found here 
between $\Nsg$ and the formation time $a_c$, 
this discrepancy might have been anticipated.

We demonstrated explicitly, in \S~\ref{sub:acc}, that deviations in the
self-similar scaling of subhalo populations with host halo mass are
due to a relative shift in the balance of power between accretion
rates and the orbital evolution timescales of subhalos.  This argument
has a natural extension.  In \S~\ref{sub:hostsub}, we took advantage
of the fact that semi-analytic approaches are not subject to the same
restrictions on sample size as direct simulation, to expound upon
correlations between host halo properties and their satellite
populations.  

We found that the host halo collapse time $\ac$ correlates strongly
with the number of satellite halos.  The sense is such that
early-forming halos have fewer satellites simply because they acquired
their satellites earlier and these satellites were therefore subject
to the destructive process of the dense, host environment for a longer
time.  This is in qualitative agreement with the recent, similar
analysis of \cite{taylor_babul04b}, who use a different definition of
halo formation time.  Formation time is strongly correlated with halo
concentration 
(\citeauthor{wechsler_etal02}; \citeauthor{zhao_etal03} \citeyear{zhao_etal03}), 
implying that satellite halo abundance is correlated
with halo concentration.  In our model, host halo 
concentration varies in
inverse proportion to $\ac$ via Eq.~\ref{eq:cofac}, so this
correlation occurs by construction, with $\Nsg \propto \cvir^{-0.86}$
for satellites selected by $\vmaxsat/\vmaxhost \ge 0.15$.

We confirmed that this correlation between halo concentration and
subhalo abundance is exhibited by host halos in simulations 
(Fig.~\ref{fig:ACNsat}).  For the $\lcdme$
simulation, we find that 
$\Nsg(\vmaxsat/\vmaxhost>0.15) \propto \cvir^{-0.98}$, 
in good agreement with the results of our model.  An
extension of this argument is that the strength of the $\ac$-$\Nsg$
(or $\cvir$-$\Nsg$) correlation should depend upon the $\vmax$
threshold used to define the satellite halo sample.  We find that at
any particular threshold of $\vmaxsat/\vmaxhost$, $\Nsg \propto
\ac^{b}$, where $b \simeq 0.25 + 4(\vmaxsat/\vmaxhost)$, for $0.08
\le \vmaxsat/\vmaxhost \le 0.5$ at redshift $z=0$.  
We find that the correlation between formation time and satellite 
abundance (or, equivalently, between host concentration and satellite 
abundance) becomes weaker at high redshift as shown in 
Fig.~\ref{fig:acnpl}.  By redshift $z=1$, the relation between 
the number of satellites and the host halo formation time is 
described by a power law index $b \simeq 2.7(\vmaxsat/\vmaxhost)$.

These results may have direct observational implications.  For
example, cluster halos of fixed mass but with below average
substructure counts are objects that acquired their substructure
earlier, allowing satellites to experience significant dynamical
evolution.  If accretion into a cluster environment strips gas from
the subhalo and effectively truncates star formation, 
this suggests that these early-forming 
clusters may have a higher fraction of red galaxies.
More directly, in a sample of clusters restricted about a narrow range
of X-ray temperatures or luminosities, cluster richness may be
expected to correlate with the fraction of red galaxies.

Similarly, central galaxy mergers would be more common in these
early-forming systems, likely causing the central galaxy to become
more luminous.  This may induce a correlation between cluster richness
and the luminosity of the brightest cluster galaxy in systems of fixed
dynamical mass.  We are tempted to identify the so-called 
{\em fossil groups} with the extreme, early-formation tail of the formation
epoch ($\ac$) distribution in our model.  The fossil groups are
systems that are characterized by X-ray luminosities comparable to
those of poor galaxy clusters, while their optical luminosities are
dominated by a single, bright, central galaxy
\citep[e.g.,][]{vikhlinin_etal99,jones_etal03}.  In this scenario,
these fossil groups would correspond to large, group-size halos that
assembled their mass very early, when dynamical timescales were short,
allowing any substructure to evolve significantly.  Any large
satellites would then be more vulnerable to a rapid merger with the
central object due to dynamical friction, leaving a luminous central
galaxy, and to severe tidal disruption, leaving a surrounding halo
bereft of luminous companions.  This interpretation is consistent with
the analyses of the radial distribution of low-luminosity satellite
galaxies in such groups \citep{mathews_etal04}.

The correlation between $\Nsg$ and $\cvir$ may be detectable
observationally.  Our results suggest that for clusters with similar
mass estimates, the optically-richer clusters should have underlying
dark matter halos with concentrations that are lower than average.  It
may be possible to detect directly trends of this kind using cluster
mass profile estimates either via X-ray analyses or gravitational
lensing.  If a trend between formation time or concentration and the
number of satellite galaxies $N_{gal}$, could be detected using
optical data, it could potentially provide an optical mass measurement
with less scatter than $N_{gal}$ or total cluster luminosity.
Specifically, a large fraction of the scatter in mass at a fixed
observed value of $N_{gal}$ in the newest generation of
optically-selected cluster catalogs is due to the
theoretically-expected scatter in $N_{gal}$ at fixed mass
(R. H. Wechsler et al., in preparation).  The aforementioned
correlation suggests that clusters at fixed $N_{gal}$ are a mix of
later-forming, low-mass clusters and earlier-forming, high-mass
clusters.  Thus, it may be possible to remove some of this scatter
with other optical observables.

Finally, we demonstrated the utility of the presented model with an
explicit calculation of correlation functions in \S~\ref{sub:app}.  We
generated a set of ten hybrid halo catalogs using the host halos from
the $\lcdme$ simulation and ten semi-analytically computed subhalo
populations for each host.  We computed the two-point correlation
function of the $\lcdme$ halos and compared it with the correlation
functions of the halos in the hybrid catalogs.  We found the
correlation functions calculated in this way to be in good agreement.
The host concentrations and the hybrid model subhalos are assigned
according to a technique that is ignorant of the environment of the
halo.  This result implies that any correlations between $\cvir$,
$\Nsg(>\vmaxsat)$, or subhalo spatial distributions with environment
are sufficiently weak that they likely do not have a measurable effect
on the correlation function.  This supports one of the basic
assumptions of halo model analysis of galaxy clustering
\citep[e.g.][]{zehavi_etal03,zehavi_etal04}, namely that the galaxy
occupation of host halos is statistically independent of environment
and depends solely upon the mass of the halo.  This also suggests that
the technique of populating a large-volume, low-resolution simulation
with semi-analytic subhalo populations is viable for studying galaxy
clustering and environment.  We pursue these issues in a forthcoming
paper in which we also study mapping galaxies onto halos using various
prescriptions in order to test several hypotheses about the physics of
galaxy formation.

%
%
%
%
%

\acknowledgments 

We thank Avishai Dekel, Q. Y. Gong, Stelios Kazantzidis, Savvas
Koushiappas, Rachel Somerville, James Taylor, and Jeremy Tinker 
for useful discussions and feedback.  We thank Neal Dalal, 
Stelios Kazantzidis, Roman Scoccimarro, Frank van den Bosch, 
David Weinberg, and Zheng Zheng for valuable comments on an 
earlier draft of this manuscript and several helpful 
suggestions.  This research made use of the NASA Astrophysics Data
System.  ARZ would like to thank the Center for Cosmology and Particle
Physics at New York University for its hospitality during several
visits while this work was in progress and the 2004 Santa Fe Cosmology
Workshop where part of this work was completed.  This work was
partially funded by The Kavli Institute for Cosmological Physics at
The University of Chicago and the National Science Foundation (NSF) 
through grant No. NSF PHY 0114422. AAB is partially supported by 
NASA (grant NAG5-11669) and NSF (grant PHY-0101738).  
AVK is supported by the NSF under grants No.  AST-0206216 and
AST-0239759, by NASA through grant NAG5-13274, and by the Kavli
Institute for Cosmological Physics at the University of Chicago.  RHW
is supported by NASA through a Hubble Fellowship awarded by the Space
Telescope Science Institute, which is operated by the Association of
Universities for Research in Astronomy, Inc., for NASA, under contract
NAS 5-26555.  The simulations used in this study were performed on the
IBM RS/6000 SP3 system at the National Energy Research Scientific
Computing Center (NERSC).

\end{document}